\documentclass[conference]{IEEEtran}
\usepackage{booktabs} 
\usepackage{enumerate}
\usepackage{graphicx}
\usepackage{multirow}
\usepackage{amsmath}
\newtheorem{definition}{Definition}[section]
\usepackage{amsfonts}
\usepackage{url}
\usepackage{subfigure}
\usepackage{xcolor}
\usepackage{bm}
\usepackage[ruled]{algorithm2e}
\usepackage{stmaryrd}

\newcommand{\ignore}[1]{}

\ifCLASSINFOpdf
\else
\fi
\begin{document}
%
\title{A Broad Learning Approach for Context-Aware Mobile Application Recommendation}

\author{\IEEEauthorblockN{Tingting Liang$^{\dag}$, Lifang He$^{\P}$, Chun-Ta Lu$^{\ddag}$, Liang Chen$^{\S}$, Philip S. Yu$^{\ddag\ast}$, Jian Wu$^{\dag}$}
\IEEEauthorblockA{$^{\dag}$ College of Computer Science \& Technology, Zhejiang University, China\\
$^{\P}$ Department of Healthcare Policy and Research, Cornell University, NY, USA\\
$^{\ddag}$ Department of Computer Science, University of Illinois at Chicago, IL, USA\\
$^{\S}$ School of Data and Computer Science,
Sun Yat-Sen University, China\\
$^{\ast}$ Institute for Data Science, Tsinghua University, Beijing, China\\
\{liangtt, wujian2000\}@zju.edu.cn, \{ lifanghescut, jasonclx\}@gmail.com, clu29@uic.edu, psyu@cs.uic.edu}
}

\maketitle

\begin{abstract}
With the rapid development of mobile apps, the availability of a large number of mobile apps in application stores brings challenge to locate appropriate apps for users. 
Providing accurate mobile app recommendation for users becomes an imperative task.
Conventional approaches mainly focus on learning users' preferences and app features to predict the user-app ratings. However, most of them did not consider the interactions among the context information of apps.
To address this issue, we propose a broad learning approach for \textbf{C}ontext-\textbf{A}ware app recommendation with \textbf{T}ensor \textbf{A}nalysis (CATA).
Specifically, we utilize a tensor-based framework to effectively integrate user's preference, app category information and multi-view features to facilitate the performance of app rating prediction. The multidimensional structure is employed to capture the hidden relationships between multiple app categories with multi-view features. We develop an efficient factorization method which applies Tucker decomposition to learn the full-order interactions within multiple categories and features. Furthermore, we employ a group $\ell_{1}-$norm regularization to learn the group-wise feature importance of each view with respect to each app category. Experiments on two real-world mobile app datasets demonstrate the effectiveness of the proposed method.



\end{abstract}


%
\IEEEpeerreviewmaketitle

\section{Introduction} \label{sec:intro}
The rapid adoption of mobile devices accelerates the proliferation of mobile apps. The number of available apps in the Google Play\footnote{Google Play: https://play.google.com/store/apps} reached 2.8 million in Mar. 2017, and there have been 2.2 million mobile apps available in the Apple App Store\footnote{Apple App Store: https://itunes.apple.com/us/genre/ios/id36?mt=8} in Jan. 2017. The surge of mobile apps with diverse functions brings not only great convenience to users but also challenges for discovering  appropriate apps. As a consequence, it becomes critical to develop effective approaches of rating prediction for recommending apps for users with accuracy.

\begin{figure*}[t]
  \centering
  \includegraphics[width=0.85\linewidth]{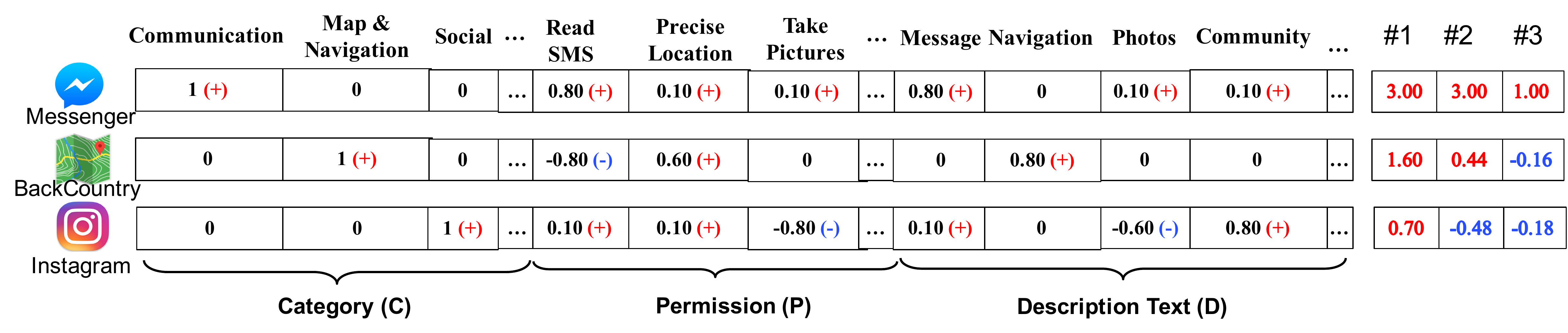}
  \caption{An example of feature interactions with different orders. The values in the column $\#1$, $\#2$, and $\#3$ represent the summation of first-order, the second-order, and the third-order interactions. $\#1=C+P+D$, $\#2=C\times P+C\times D+P\times D$, $\#3=C\times P\times D$.}\label{fig:intro}
\end{figure*}

There are some recent studies about the mobile apps recommendation, most of which leverage features of apps or users \cite{karatzoglou2012climbing} \cite{liu2015personalized} \cite{zhu2014mobile}. Karatzoglou et al. \cite{karatzoglou2012climbing} proposed a collaborative filtering method for app recommendation by incorporating some contextual features like location, time of day, etc. Liu et al. \cite{liu2015personalized} proposed to incorporate both app functionality and user privacy preference as features and capture the trade-off between them for app recommendation. 
Most of the previous works only tried one kind of feature or a simple combination of multiple features and did not consider the complex interactions between those features. Currently there exist many works exploiting multiple views of features in the tasks like recommendation, clustering, etc \cite{lu2016item, shao2015clustering, shao2016online, shao2015multiple}. In the scenario of app recommendation, the interactions between different views of features are quite important as different views can provide complementary information. 
For example, assume we have obtained the latent representations for each app from three aspects, i.e., categories, permissions and description text, as shown in Fig. \ref{fig:intro}. BackCountry Navigator is an app categorized as \emph{Maps\&Navigation} and it is mainly used for outdoor navigation which can be inferred from the description text. The permission of getting users' precise location is acceptable (\emph{i.e.,} a positive value), while the permission of reading SMS is abnormal (\emph{i.e.,} a negtive value). It can be found that only the third-order interaction provides a negative result reflecting the unreasonable permission for app function. The category of Instagram merely shows its function for social interaction (\emph{i.e.,} a positive value) and neglects the function of sharing photos (\emph{i.e.,} a negtive value). Through the interactions between multiple views, complementary information is provided to show a more sufficient understanding about the app.
Obviously, the comprehensive consideration of the features from multiple views would be more insightful on understanding app information and user preference. 

To figure out the latent correlations among the context information of apps, we conduct an empirical analysis on the dataset collected from Google Play and  discover some important characteristics of mobile apps. For apps in different categories, users' download behaviors are different. Within some categories like \emph{Maps \& Navigation} and \emph{Weather}, users might only download one or two app for a long time use. However, for some categories like \emph{Entertainment}, users are more likely to download many apps in the same category. Generally, the different download behaviors happen because users will consider different reasons (\emph{e.g,} functions, interface, permissions) to decide whether to download apps for different categories. It can be inferred that users would focus on multiple views of features with different importance for apps in different categories.
The analysis of the feature-level correlation between categories shows the similarities between different categories are lower, which implies that the significances of features within a specific view are different for different categories. The category diversities of apps rated by different users are distinct. Some users prefer to download apps from various categories even though the amount of the downloaded apps is small. 
Based on the analysis, we consider to fuse the user preference, the category information, the features of multiple views, and the complex interactions among them to generate a context-aware category specific app rating prediction model.



In this paper, we propose a broad learning approach for \textbf{C}ontext-\textbf{A}ware app recommendation with \textbf{T}ensor \textbf{A}nalysis (CATA). Specifically, we integrate the interactions among the multiple categories and multiple views of features into a tensor structure through the tensor product of the corresponding feature spaces. The interactions with different orders can fully reflect the complementary relationships, and we use them to predict the user ratings on apps. To effectively learn the full-order interactions \footnote{Full-order interactions range from the first-order interactions (i.e., single-view features in each category) to the highest-order interactions (i.e., all combinations of features from multiple views and from different categories).} without physically building a tensor, we further develop an efficient factorization method which employs Tucker decomposition. The Tucker decomposition is applied to factorize the interaction parameters for each order, which can make accurate parameter estimation under sparsity and avoid overfitting. Moreover, we introduce the group $\ell_{1}$-norm regularization for the global-specific weight matrix to further improve the proposed model.

The main contributions of this paper are as follows:
\begin{itemize}
  \item We propose a context-aware recommendation approach for mobile apps called CATA that models the interactions with different orders among the multiple categories and multiple views of features as a tensor structure. 
  \item To effectively learn the hidden relationships among the different views of the context information of apps, Tucker decomposition is adopted to factorize the interaction parameters such that the principal components of the latent representations can be retained. 
  \item Empirical studies based on two real world datasets demonstrate the effectiveness of the proposed context-aware recommendation approach.
\end{itemize}


\section{Data Analysis} \label{sec:dataset}
In this section, we first describe the datasets used for the analysis and experiments. We then provide the statistical characteristics of the employed datasets.
\subsection{Data Description}

\begin{itemize}
  \item \textbf{Google Play:} We crawled app's meta data (e.g., name, category, permissions, description) and user review ratings from its description page in Google Play. We filter users and apps with less than 5 ratings. Each rating record in this dataset is represented in three views, \emph{i.e.,} users, permissions and text. The user view consists of binary feature vectors for user ids which means there is only one non-zero feature in the user view for each rating record. The TF-IDF vector representations of the app permissions and description texts are used as the permission and text view, respectively. \\
  
  \item \textbf{Apple's App Store:} The dataset is offered by \cite{zhu2015discovery}\cite{zhu2015popularity} and consists of the apps in the ``Top Free 300'' and ``Top Paid 300'' leaderboards from Feb. 2010 to Sep. 2012, and the related user ratings and review information. As the dataset lacks of the classification information, we use \emph{Free} and \emph{Paid} as two categories, and we remove users and apps with less than 10 ratings. Each rating record in this dataset has two views, \emph{i.e.,} users and text. The user view are constructed using the same way as in the Google Play dataset. The TF-IDF vector representations of the review texts of apps are used as the text view.
\end{itemize}

Table \ref{tab:data} shows the basic statistics of the two employed datasets.

\begin{table}[t]\scriptsize
  \centering
  \caption{Statistics of the datasets}
    \begin{tabular}{l|c|c|c|c|c}
    \hline
    Dataset & \#App & \#User & \#Feature & \#Category & \#Rating \\
\hline    
    \multirow{2}[2]{*}{Google Play} & \multirow{2}[2]{*}{5460} & \multirow{2}[2]{*}{7165}& Text (2574) & \multirow{2}[2]{*}{45} & \multirow{2}[2]{*}{67504} \\
          &       &  &    Permissions (84)   &  \\
    \hline
    \multirow{1}[2]{*}{Apple App Store} & \multirow{1}[2]{*}{2643} & \multirow{1}[2]{*}{4010} & \multirow{1}[2]{*}{Text (1592)} & \multirow{1}[2]{*}{2} & \multirow{1}[2]{*}{74764} \\
    \bottomrule
    \end{tabular}%
  \label{tab:data}%
\end{table}%

\subsection{Characteristics of Google Play dataset}

\begin{figure*}[t]
\centering
\subfigure[Percent of users with more than 2 apps ]{\includegraphics[width=5.5cm]{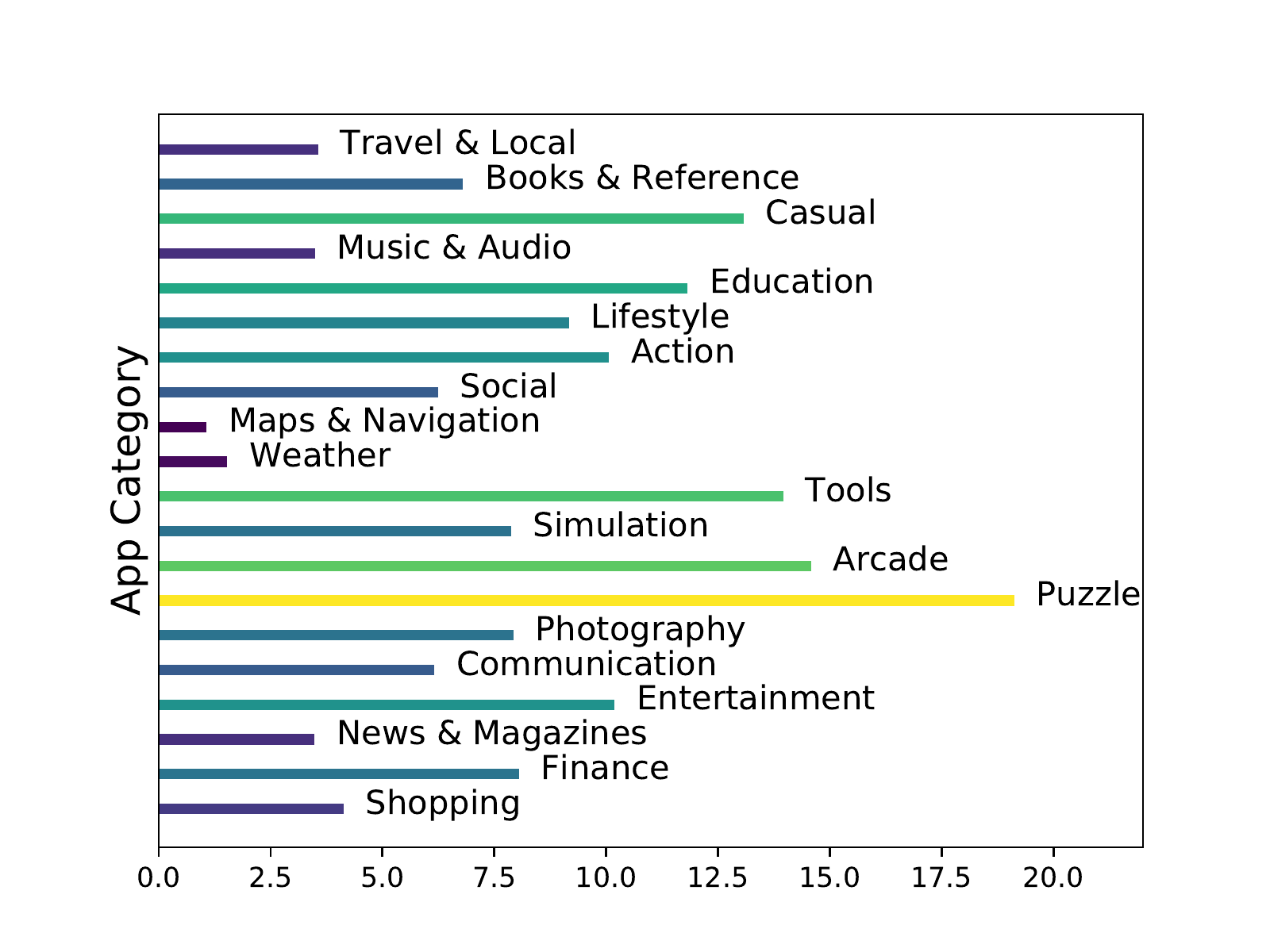}}
\subfigure[Feature similarity between categories]{\includegraphics[width=5.5cm]{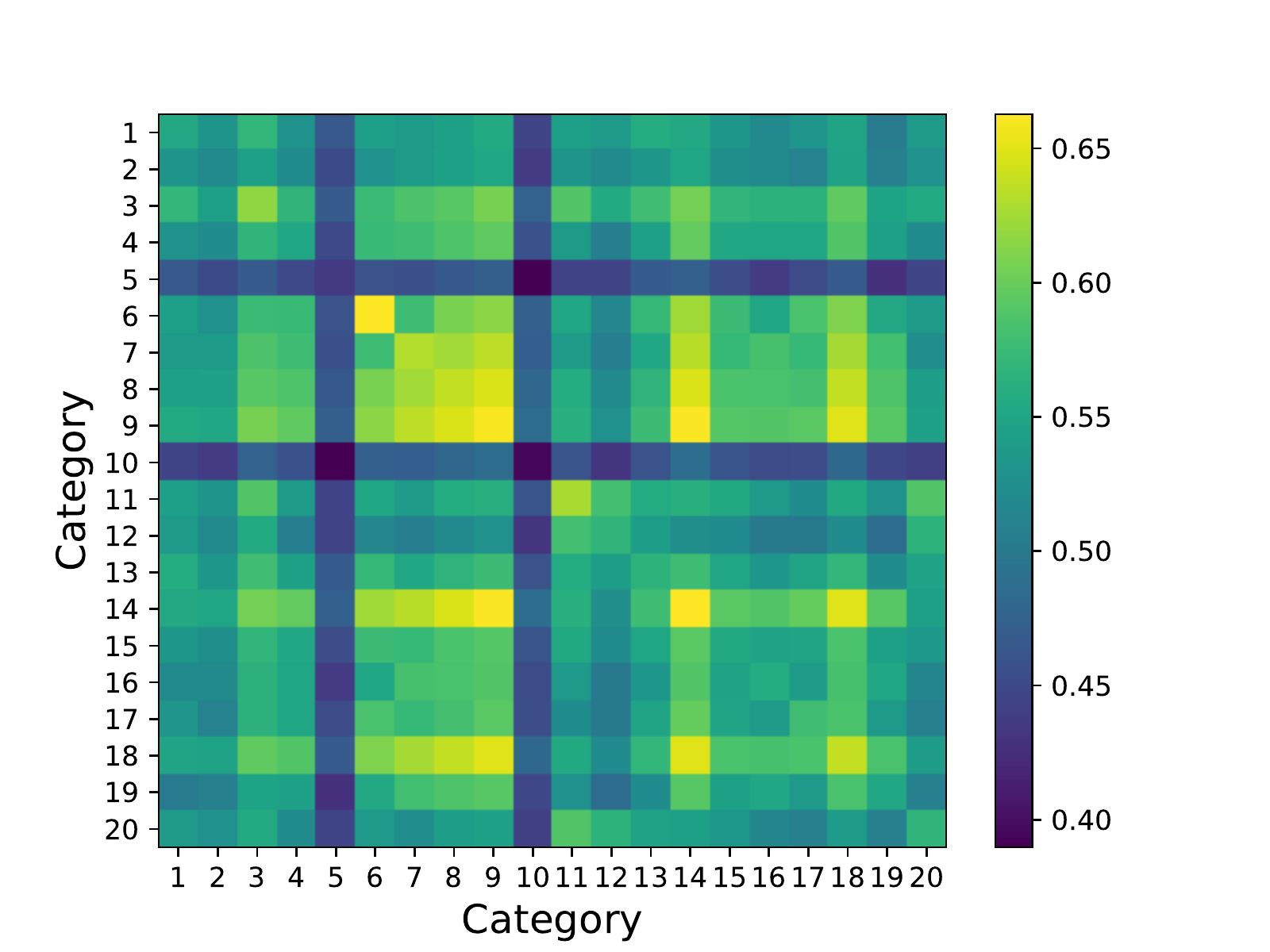}}
\subfigure[Category diversity of each user]{\includegraphics[width=5.5cm]{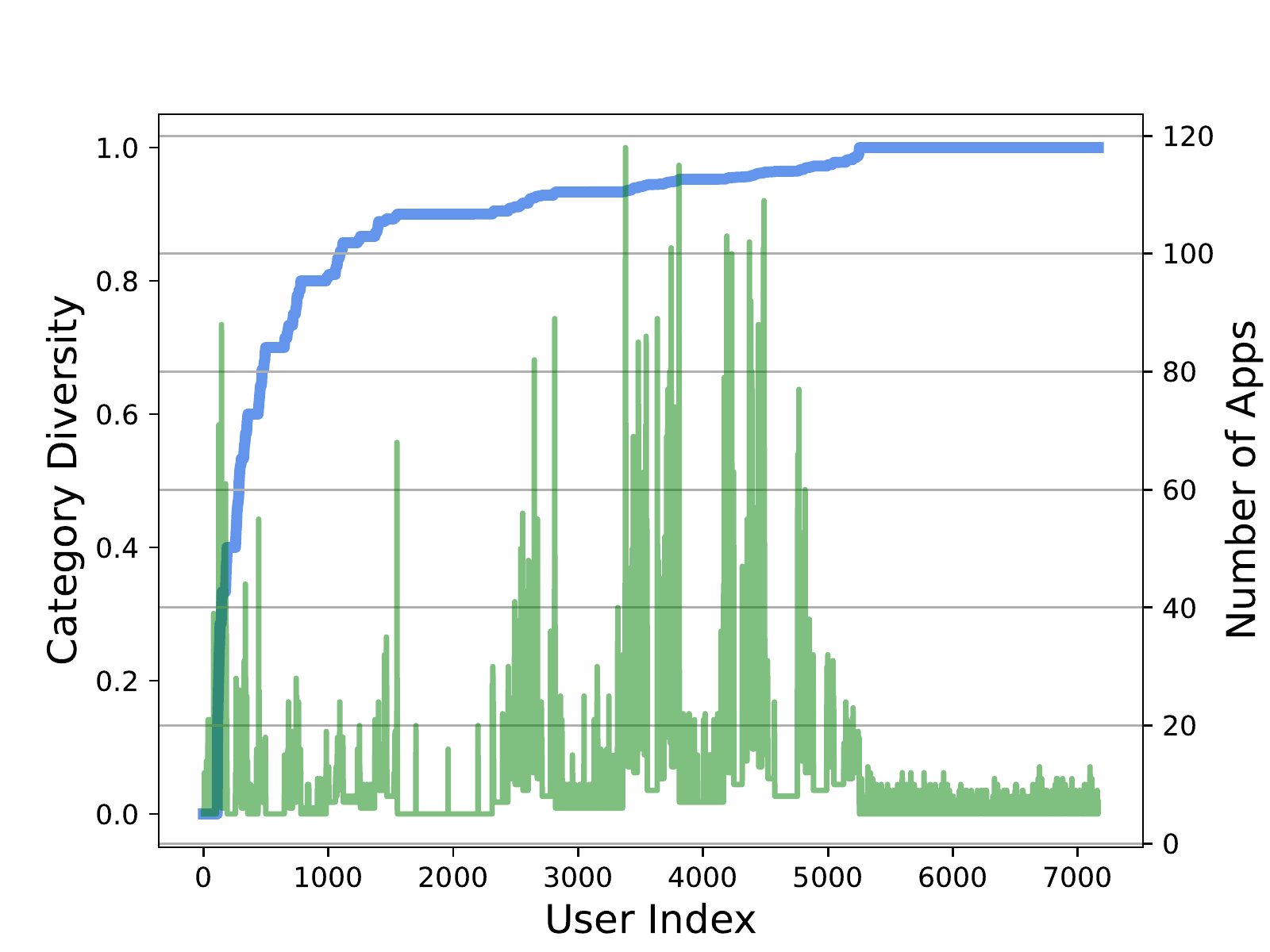}}
\caption{Characteristics of Google Play dataset. The blue curve and green curve in (c) represent category diversity and app number, respectively.}\label{fig:data_analysis}
\end{figure*}
As the Google Play dataset has richer category and feature information, we focus on the analysis for it.

We calculate the proportions of the users who downloaded more than 2 apps for each category. Due to the space limitation, Fig. \ref{fig:data_analysis}(a) reports the results of 20 randomly selected app categories. The lower proportion for a category is, the more users only download one or two apps in that category. Generally, apps in the categories with a low proportion can be used for a long time. Taking category \emph{Maps \& Navigation} in which only about 1.07\% users downloaded more than 2 apps as an example,  users usually only download one or two apps (\emph{e.g.,} \emph{Google Map},\emph{ Baidu Map} ) in this category as these apps are sufficient to use. The other categories having the low proportions in Fig \ref{fig:data_analysis}(a) are \emph{Weather}, \emph{Travel\& Local}, etc. The categories with high proportions are more general and the apps in them are with more varieties, like \emph{Tools}, \emph{Puzzle}, \emph{Arcade}, etc. 
For apps in these categories, users might consider different reasons to decide whether to download the apps. For example, users mainly consider if the functions of the apps in category \emph{Tools} will meet their demands. But for apps in category \emph{Maps \& Navigation}, compared to the functions, users pay more attention to the features like interface or permissions as they have clearly known the functions. We can learn that users would focus on multiple views of features with different significances for apps in different categories. 

After the investigation of relationship between multiple views and categories, we explore the features within a specific view. To investigate the feature-level correlation between categories with respect to a certain view, we calculate the feature-based similarities between each pair of apps from any two categories. Figure \ref{fig:data_analysis}(b) shows the similarities generated based on app permission feature between any two categories. The category indexes are sorted by the order in Fig. \ref{fig:data_analysis}(a) (\emph{i.e.,} \#1 is \emph{Shopping}). It can be observed that the similarities between two different categories are generally lower than those between the same categories. That is, for apps of different categories, the significance of features within a specific view are distinguishing. 



To investigate the relationship between users and categories, we apply the diversity metric widely used for the evaluation of recommender systems \cite{ricci2011introduction} to evaluate the category diversity. The category diversity is calculated by $Div(u) = 1 - \frac{\sum_{i, j\in A(u),i\neq j}{s(i,j)}}{\frac{1}{2}|A(u)|(|A(u)|-1)}$,
where $A(u)$ is the set of apps rated by user $u$. $s(i,j) = 1$ if app $i$ and $j$ belong to the same category, otherwise, $s(i,j) = 0$. Figure \ref{fig:data_analysis}(c) shows the category diversity and the number of apps for each user. 
The green curve presents the number of apps rated by users, and the blue curve is the category diversity of the apps rated by users. 
The user indexes on the $x$-axis are sorted by the values of category diversity in an ascending order. The left $y$-axis shows the value of category diversity and the right $y$-axis represents the number of apps. 
It can be found that some users have interactions with many types of apps even though the numbers of apps rated by them are very small while some users rate many apps with few categories. Different users have interactions with categories with different diversities. As discussed above, the importance of features from multiple views and features within a specific view is distinct for different categories. Therefore, for each user, it is critical to model his preference on an app considering the category information and the corresponding relationships with features of multiple views.



Based on the analysis of the relationships among app category, app feature, and user, it requires a recommendation model which can integrate the interactions among the multiple categories, multiple views of features, and users.


\section{Preliminaries} \label{sec:preliminary}
In this work, we intend to predict ratings for mobile applications by a tensor-based approach. 
Before that, we introduce some related concepts and notation in tensor algebra that will be used throughout the paper, and then provide the problem formulation of app rating prediction.


\subsection{Tensor Concepts and Notation} 
A tensor is a multi-dimensional array which generalizes matrix representation. Each dimension in tensor is called \emph{mode} or \emph{way}. Following prevailing convention, tensors are represented by calligraphic letters, matrices by boldface uppercase letters, vectors by boldfaced lowercase letters, and scalars by lowercase letters. An element of a vector $\mathbf{x}$, a matrix $\mathbf{X}$, or a tensor $\mathcal{X}$ is represented by $x_i$, $x_{i,j}$, $x_{i,j,k}$, etc., depending on the number of modes. All vectors are column vectors unless otherwise specified. For an arbitrary matrix $\mathbf{X} \in \mathbb{R}^{I \times J}$, its $i$-th row and $j$-th column vector are represented by $\mathbf{x}^{i}$ and $\mathbf{x}_{j}$, respectively. The outer product of $N$ vectors $\mathbf{x}^{(n)} \in \mathbb{R}^{I_n}$ for all $n \in [1:N]$ is an $N$-th-order tensor and defined elementwise as $\big(\mathbf{x}^{(1)} \circ \cdots \circ \mathbf{x}^{(N)}\big)_{i_1, \dots, i_N} = x^{(1)}_{i_1}  \cdots x^{(N)}_{i_N}$ for $i_n \in [1:I_n]$. The inner product of two tensors $\mathcal{X}, \mathcal{Y}\in \mathbb{R}^{I_1\times\cdots \times I_N}$ is defined as $\big\langle\mathcal{X}, \mathcal{Y} \big \rangle = \sum_{i_1=1}^{I_1}\cdots \sum_{i_N=1}^{I_N}x_{i_1, \dots, i_N}y_{i_1,\dots, i_N}$.
In particular, for $\mathcal{X} = \mathbf{x}^{(1)}\circ\cdots\circ\mathbf{x}^{(N)}$ and $\mathcal{Y} = \mathbf{y}^{(1)}\circ\cdots\circ\mathbf{y}^{(N)}$, it holds that 
\begin{equation}\label{eq:xy}
\big\langle\mathcal{X}, \mathcal{Y} \big \rangle = \prod_{n=1}^{N}\big\langle\mathbf{x}^{(n)},\mathbf{y}^{(n)}\big\rangle = \prod_{n=1}^{N}\mathbf{x}^{(n)^{T}}\mathbf{y}^{(n)}.
\end{equation}

Definitions of Kronecker product, Khatri$-$Rao product, mode-$n$ product, and Tucker decomposition are given below, which will be applied to build the proposed model.



\begin{definition}[Kronecker Product]
The \emph{Kronecker product} of matrices $\mathbf{X}\in\mathbb{R}^{I\times J}$ and $\mathbf{Y}\in\mathbb{R}^{K\times L}$ is denoted by $\mathbf{X}\otimes\mathbf{Y}$. The result is a matrix of size $(IK)\times(JL)$ and defined by
\begin{small}
\begin{equation}
\begin{split}    
\mathbf{X}\otimes\mathbf{Y}
&=
\begin{bmatrix}
x_{1,1}\mathbf{Y} &x_{1,2}\mathbf{Y}  & \cdots &x_{1,J}\mathbf{Y} \\ 
 \vdots &  \vdots & \ddots & \vdots  \\ 
x_{I,1}\mathbf{Y} &x_{I,2}\mathbf{Y}  & \cdots &x_{I,J}\mathbf{Y}  
\end{bmatrix}\\
&=\begin{bmatrix}
\mathbf{x}_{1}\otimes\mathbf{y}_{1} & \mathbf{x}_{1}\otimes\mathbf{y}_{2} & \cdots & \mathbf{x}_{J}\otimes\mathbf{y}_{L-1} & \mathbf{x}_{J}\otimes\mathbf{y}_{L}
\end{bmatrix}.
\end{split}
\end{equation}
\end{small}
\end{definition}

\begin{definition}[Khatri$-$Rao Product]
The \emph{Khatri$-$Rao product} of matrices $\mathbf{X}\in\mathbb{R}^{I\times K}$ and $\mathbf{Y}\in\mathbb{R}^{J\times K}$ is denoted by $\mathbf{X}\odot\mathbf{Y}$. The result is a matrix of size $(IJ)\times K$ and defined by
\begin{equation}
\mathbf{X}\odot\mathbf{Y}=
\begin{bmatrix}
\mathbf{x}_{1}\otimes\mathbf{y}_{1} & \mathbf{x}_{2}\otimes\mathbf{y}_{2} & \cdots & \mathbf{x}_{K}\otimes\mathbf{y}_{K}
\end{bmatrix}.
\end{equation}
\end{definition}

The Khatri$-$Rao product is the ``matching columnwise'' Kronecker product.

\begin{definition}[$n$-mode Product]
The \emph{$n$-mode product} of a tensor $\mathcal{X}\in \mathbb{R}^{I_{1}\times\cdots \times I_N}$ with a matrix $\mathbf{U}\in \mathbb{R}^{J\times I_n}$ denoted by $\mathcal{X} \times_{n}\mathbf{U}$ is defined as
\begin{equation}
(\mathcal{X}\times_{n}\mathbf{U})_{i_1,\dots, i_{n-1},j,i_{n+1},\dots, i_N} = \sum_{i_n=1}^{I_n}x_{i_1, \dots, i_N}u_{j,i_n}.
\end{equation}
\end{definition}

\begin{figure}[t]
  \centering
  \includegraphics[width=8cm]{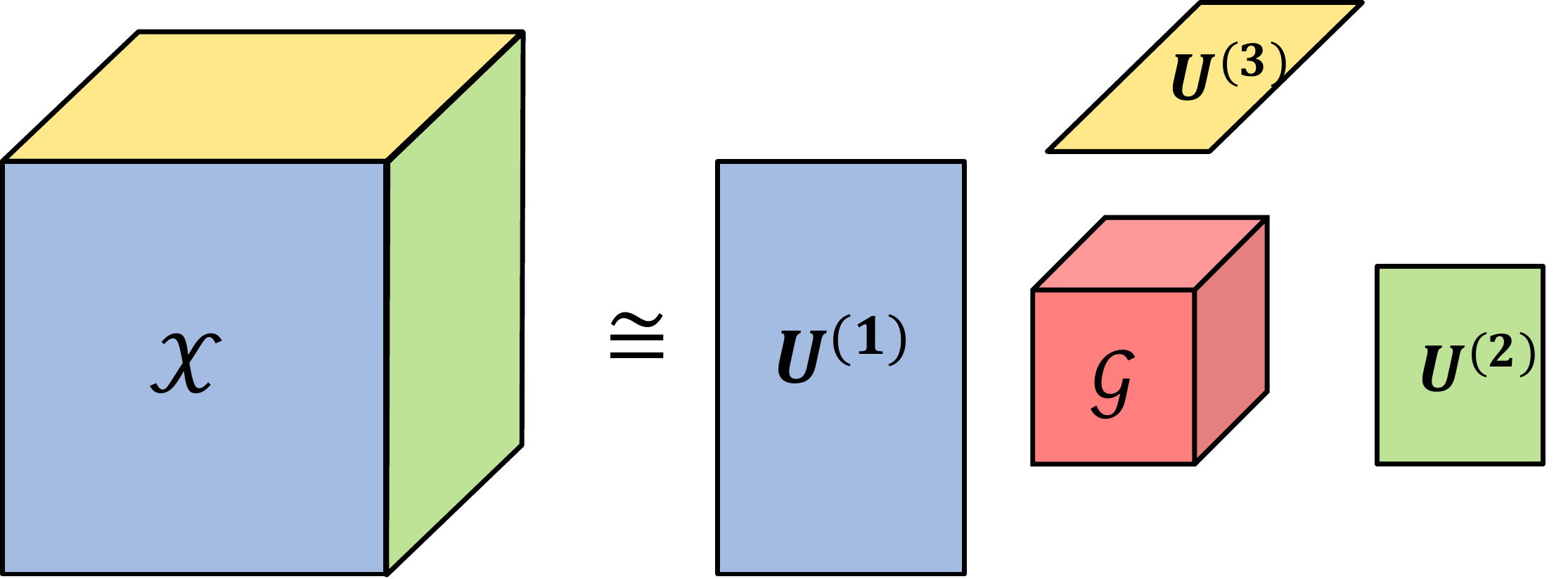}
  \caption{Tucker decomposition of a third-order weight tensor.}\label{fig:tucker}
\end{figure}

Figure \ref{fig:tucker} visualizes the Tucker decomposition of a third-order tensor and table \ref{tab:notation} summarizes the main notations for easy referencing.

\begin{definition}[Tucker Decomposition]
For a general tensor $\mathcal{X}\in \mathbb{R}^{I_1\times\cdots\times I_N}$, its Tucker decomposition is defined as 
\begin{equation}
\begin{split}
\mathcal{X} & \approx \mathcal{G}\times_{1}\mathbf{U}^{(1)}\times_{2}\cdots\times_{N}\mathbf{U}^{(N)}\\
& = \sum_{r_1=1}^{R_1}\cdots\sum_{r_N=1}^{R_N}g_{r_1,\dots, r_N}\mathbf{u}_{r_1}^{(1)}\circ\cdots\circ\mathbf{u}_{r_N}^{(N)} \\
&= \llbracket \mathcal{G}; \mathbf{U}^{(1)},\cdots, \mathbf{U}^{(N)}\rrbracket,
\end{split}
\end{equation}
where $\mathbf{U}^{(n)}\in\mathbb{R}^{I_{n}\times R_{n}}$ are the factor matrices and can be thought of as the principal components in each mode. $\mathcal{G}\in \mathbb{R}^{R_{1}\times\cdots\times R_{N}}$ is called the \emph{core tensor}. $\llbracket\cdot \rrbracket$ is used for short-hand notation.
\end{definition}

\begin{table}
\small
\caption{List of basic symbols.}
\label{tab:notation}
\begin{tabular}{ll}
\toprule
Symbol & Definition and description\\
\midrule
$x$ & each lowercase letter represents a scale\\
$\mathbf{x}$ & each boldface lowercase letter represents a vector\\
$\mathbf{X}$ & each boldface capital letter represents a matrix\\
$\mathcal{X}$ & each calligraphic letter represents a tensor, set or space\\
$[1:N]$ & a set of integers in the range of $1$ to $N$ inclusively. \\
$\left\langle\cdot,\cdot\right\rangle$ & denotes inner product\\
$\circ$ & denotes outer product\\
$\otimes$ & denotes Kronecker product\\
$\odot$ & denotes Khatri$-$Rao product\\
$\times_n$ & denotes $n$-mode product\\
$\left\|\cdot\right\|_{F}$ & denotes Frobenius norm of vector, matrix or tensor\\
\bottomrule
\end{tabular}
\end{table}

\subsection{Problem Formulation}
Suppose that  the scenario of app rating prediction includes a user set $\mathcal{U}$ and mobile app set $\mathcal{A}$. The numbers of app categories and feature views are $C$ and $V$. Let $N_{c}$ be the number of the rating records in the category $c\in [1:C]$, then the total number of rating records is $N = \sum_{c}N_{c}$. Let $I_{v}$ be the dimensionality of the feature view $v\in [1:V]$ and $I = \sum_{v}I_{v}$.

In this paper, we construct a multi-dimensional tensor to discover the latent interactions among the category information and multi-view features.
Each rating record in category $c$ can be represented in $V$ different views, i.e., $\mathbf{x}_{c}^{T} = (\mathbf{x}_{c}^{(1)^{T}}, \dots, \mathbf{x}_{c}^{(V)^{T}})$, where $\mathbf{x}_{c}^{(v)}\in \mathbb{R}^{I_{v}}$ and $\mathbf{x}_{c}\in \mathbb{R}^{I}$. Generally, a rating record involves a user, an app, and different types of characteristics of the app. Given a training set of rating records $\mathcal{D} = \{ (\mathbf{X}_{c}^{(1)}, \dots , \mathbf{X}_{c}^{(V)} , \mathbf{y}_{c}) | c\in [1:C]\}$, where $\mathbf{X}_{c}^{(v)}\in \mathbb{R}^{I_v\times N_t}$ is the feature matrix in the $c$-th category for $v$-th view and $\mathbf{y}_{c}$ is the vector of the rating values of those apps in the $c$-th category. Our goal is to find a predictive function $f_{c} : \mathcal{X}_{c} \rightarrow \mathcal{Y}_{c}$ for each category that can minimize the expected loss and provide accurate predicted ratings. The regularized objective function to be minimized can be formulated as: 
\begin{equation}
\mathcal{H}(\{f_{c}\}_{c=1}^{C}) = \sum_{c=1}^{C}\big( \mathcal{L}_{c}\big(f_{c}(\{\mathbf{X}_{c}^{(v)}\}),\mathbf{y}_{c}\big )\big ) + \lambda\Omega ,
\end{equation}
where $\mathcal{L}_{c}$ is the empirical loss in the $c$-category. $\Omega$ is the regularization term and $\lambda >0$ is the regularization parameter. $\mathcal{L}_{c}$ can be rewritten as the average square error of each instance.
\begin{equation}
\begin{split}
\mathcal{L}_{c}\big(f_{c}(\{\mathbf{X}_{c}^{(v)}\}),\mathbf{y}_{c}\big ) 
&= \frac{1}{N_{c}}\sum_{n=1}^{N_c}\ell\big(f_{c}(\{\mathbf{x}_{c,n}^{(v)}\}),\mathbf{y}_{c,n}\big )\\
&= \frac{1}{N_{c}}\sum_{n=1}^{N_c}\big(f_{c}(\{\mathbf{x}_{c,n}^{(v)}\}) - \mathbf{y}_{c,n}\big)^{2}.
\end{split}
\end{equation}

\section{Proposed Method} \label{sec:method}
In this section, we first introduce the context-aware recommendation approach based on tensor analysis (CATA). Then we discuss how to employ Tucker decomposition to learn the proposed model without physically building the tensor. 


\subsection{Model for App Rating Prediction}
We derive the proposed model from the basic framework of linear analysis. Given a vector of an app rating record $\mathbf{x}\in \mathbb{R}^{I}$, the basic linear model for the $c$-th category is written as
\begin{equation}
f_{c}(\mathbf{x}) = \sum_{i=1}^{I}w_{c,i}x_{i} + w_{c,0} = \mathbf{x}^\mathrm{T}\mathbf{w}_{c} + w_{c,0},
\label{eq:linear_base}
\end{equation}
where $\mathbf{w}_{c}\in\mathbb{R}^{I}$ is the weight vector for the $c$-th category, and $w_{c,0}$ is the bias factor for adjusting the threshold of the $c$-th category label assignment.

Let $\mathbf{z} = [1; \mathbf{x}] \in \mathbb{R}^{1+I}$ and $\mathbf{w}_c = [w_{c, 0}; \mathbf{w}_c] \in \mathbb{R}^{(1+I)}$, then the bias factor $w_{c,0}$ can be absorbed to $\mathbf{w}_c$ (see \cite{shalev2014understanding}). Eq.~(\ref{eq:linear_base}) can thus be rewritten as follows:
\begin{equation}
f_{c}(\mathbf{x}) = \mathbf{z}^\mathrm{T}\mathbf{w}_{c}.
\label{eq:linear_base1}
\end{equation}
Let $\mathbf{W} \in \mathbb{R}^{(1+I) \times C}$ denote the weight matrix to be learned, whose columns are the vector $\mathbf{w}_{c}$. In order to jointly learn multiple linear models for $C$ categories, we introduce a category indicator vector denoted by $\mathbf{e}_{c}\in \mathbb{R}^{C}$ to model the second-order interactions between input features and categories. The indicator vector $\mathbf{e}_{c}$ is defined as
\begin{equation*}
\mathbf{e}_{c} = [\underbrace{0,\cdots,0}_{c-1},1,0,\cdots,0]^\mathrm{T}.
\end{equation*}
Then Eq.~(\ref{eq:linear_base1}) can be rewritten as
\begin{equation}\label{eq:category}
f_{c}(\mathbf{x}) = \mathbf{z}^{T}\mathbf{w}_{c} = \mathbf{z}^\mathrm{T}\mathbf{W}\mathbf{e}_{c} = \left\langle\mathbf{W}, \mathbf{z}\circ\mathbf{e}_{c}\right\rangle.
\end{equation}
Note that the outer product is used to compute intersections between input features and categories, which consists in the product of all combinations of the variables that define each domain. This data fusion technique provides a good framework to introduce multiple features. When each object is associated with multi-view features, by means of the outer product we can easily extend the above Eq. (\ref{eq:category}) to the multi-view case and provide a consensus formulation.

Suppose that the given rating records are composed by features of $V$ views (denoted as $\{\mathbf{x}^{(v)}\}, v \in [1:V]$), we can extend Eq.~(\ref{eq:category}) to model the full-order interactions between multi-view features and categories as:
\begin{equation}\label{eq:full-order1}
f_{c}(\{\mathbf{x}^{(v)}\}) =  \big\langle\mathcal{W}, \mathbf{z}^{(1)}\circ\cdots\circ\mathbf{z}^{(V)}\circ\mathbf{e}_{c}\big\rangle,
\end{equation}
or element-wise as
\begin{small}
\begin{equation}\label{eq:full-order2}
f_{c}(\{\mathbf{x}^{(v)}\}) = \sum_{s=1}^{C}\sum_{i_1=0}^{I_1}\cdots\sum_{i_V=0}^{I_V}w_{i_1,\dots,i_V,s}\big(e_{c,s}\prod_{v=1}^{V}z_{i_v}^{(v)}\big). 
\end{equation}
\end{small}
Where $\mathbf{z}^{(v)} = [1; \mathbf{x}^{(v)}] \in \mathbb{R}^{(1+I_v)}$ is the input data vector, and $\mathcal{W} = \{w_{i_1,\dots,i_V,s}\}\in\mathbb{R}^{(1+I_1)\times\cdots\times(1+I_V)\times C}$ is the weight tensor to be learned, wherein $w_{0,\dots,0}$ is the global bias, and $w_{i_1,\dots, i_V, s}$ with some indexes satisfying $i_{v}=0$ encodes lower-order interactions between views whose $i_{v'}>0$.

\begin{figure}[t]
  \centering
  \includegraphics[width=9cm]{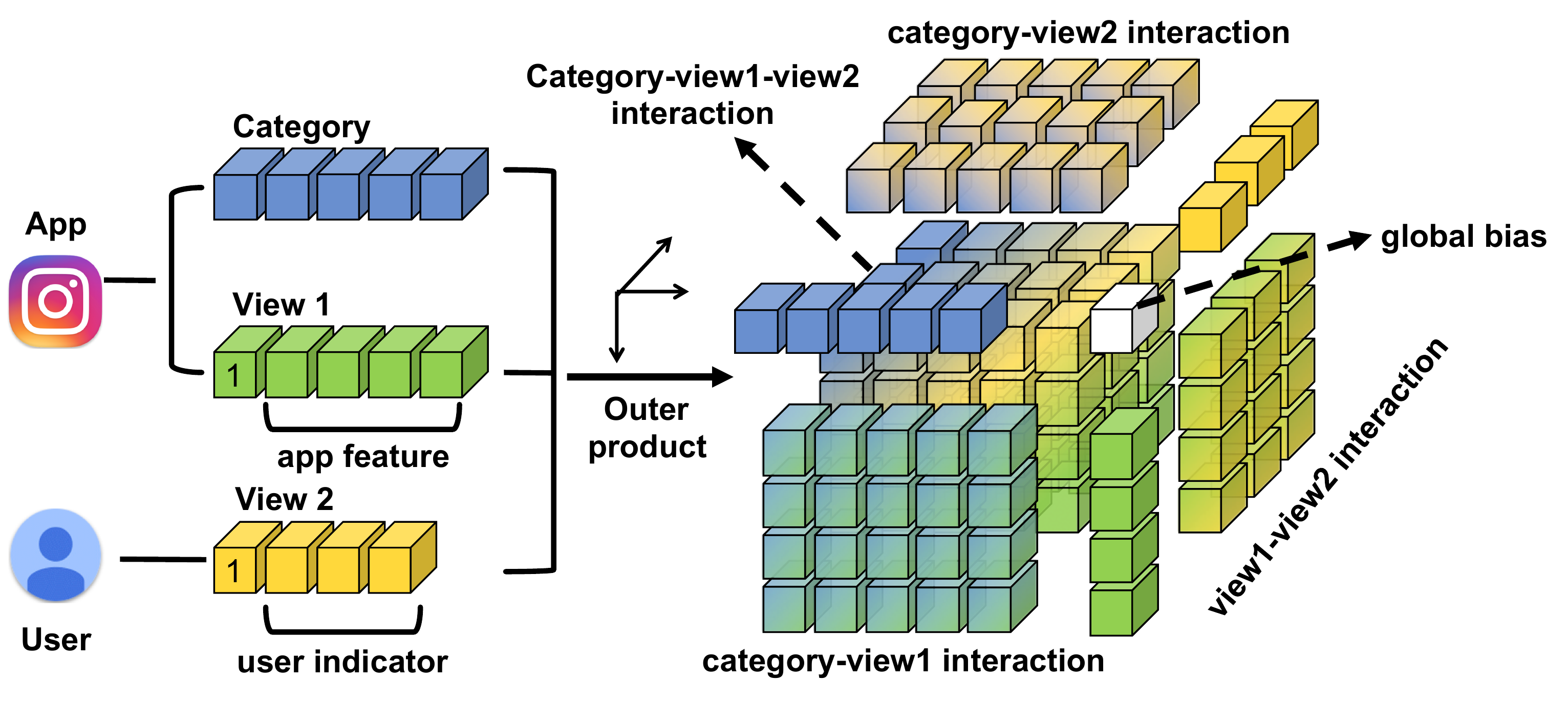}
  \caption{The context information of each rating record are represented by multiple vectors and combined by the outer product to generate the full-order interactions. The full-order interactions between multiple views and categories are modeled in a tensor structure.}\label{fig:tensor_form}
\end{figure}

In such a manner, the full-order interactions between multiple views and categories are embedded within the tensor structure, as shown in Fig. \ref{fig:tensor_form}. However, one drawback might be generated from the model is that not all the categories are fit to the constructed feature tensor and those interactions will be redundant information. Thus, we consider to build a rating predictive function based on both the full-order feature interaction space and the original feature spaces. 

Let $\mathcal{Z}_{c} = \mathbf{z}^{(1)}\circ\cdots\circ\mathbf{z}^{(v)}\circ\mathbf{e}_{c}\in \mathbb{R}^{(1+I_1)\times\cdots\times(1+I_V)\times C} $ be the full-order tensor, and $\mathbf{x} = [\mathbf{x}^{(1)};\dots;\mathbf{x}^{(V)}]\in\mathbb{R}^{I}$ be the feature vector concatenated by multiple views. We formulate our CATA model as follows:
\begin{equation}\label{eq:pre_func}
f_{c}(\{\mathbf{x}^{(v)}\}) = \left \langle \mathcal{W},\mathcal{Z}_{c} \right\rangle + \mathbf{x}^{T}\mathbf{d}_{c},
\end{equation}
where $\mathbf{d}_{c}\in\mathbb{R}^{I}$ is the category-specific weight vector. For convenience in the following discussion, we denote $\mathbf{D} = [\mathbf{d}_{1}, \cdots, \mathbf{d}_{C}] \in\mathbb{R}^{I\times C}$.



\subsection{Model Inference}
The number of parameters to be estimated in Eq. (\ref{eq:pre_func}) is $C(\prod_{v=1}^{V}(1+I_{v}) + I)$, which makes it infeasible to directly learning the model. Therefore, we assume that the weight tensor $\mathcal{W}$ can be factorized by Tucker decomposition as
\begin{equation}
\mathcal{W}= \llbracket\mathcal{G};\bm{\Phi} ,\bm{\Theta}^{(1)},...,\bm{\Theta}^{(V)}\rrbracket,
\end{equation}
where $\mathcal{G}\in \mathbb{R}^{R_{0}\times R_{1}\times ...\times R_{V}}$ is called the core tensor and its entries show the level of interaction between the different components, $\bm{\Theta}^{(v)}\in \mathbb{R}^{(1+I_{v})\times R_{v}}$ is the shared structure matrix for the $v$-th view, and $\bm{\Phi}\in \mathbb{R}^{C\times R_{0}}$ is the category specific weight matrix.


Then we can transform Eq. (\ref{eq:full-order2}) into
\begin{footnotesize}
\begin{align}
& \left \langle \mathcal{W},\mathcal{Z}_{c}  \right \rangle = \sum_{s=1}^{C}\sum_{i_1=0}^{I_1}\cdots\sum_{i_V=0}^{I_V}w_{i_1,\dots,i_V,s}\big(e_{c,s}\prod_{v=1}^{V}z_{i_v}^{(v)}\big) \nonumber \\
& = \sum_{s=1}^{C}\sum_{i_{1}=0}^{I_{1}}\cdots \sum_{i_{V}=0}^{I_{V}} \big(\sum_{r_{0}=1}^{R_{0}}\cdots \sum_{r_{V}=1}^{R_{V}}g_{r_{0},\dots, r_{V}}\phi_{s,r_{0}}\prod_{v=1}^{V}\theta_{i_{v},r_{v}}^{(v)}\big)\big(e_{c,s}\prod_{v=1}^{V}z_{i_{v}}^{(v)}\big) \nonumber \\
& = \sum_{r_{0}=1}^{R_{0}}\cdots \sum_{r_{V}=1}^{R_{V}}g_{r_{0},\dots, r_{V}}\big(\sum_{s=1}^{C}\phi_{s,r_{0}}e_{c,s}\big) 
\sum_{i_{1}=0}^{I_{1}}\cdots \sum_{i_{V}=0}^{I_{V}}\big( \prod_{v=1}^{V}\theta_{i_{v},r_{v}}^{(v)}z_{i_{v}}^{(v)}\big) \nonumber\\
& = \sum_{r_{0}=1}^{R_{0}}\cdots \sum_{r_{V}=1}^{R_{V}}g_{r_{0},\dots, r_{V}}\big\langle \bm{\theta}_{r_{1}}^{(1)}\circ\cdots\circ \bm{\theta}_{r_{V}}^{(V)}\circ \bm{\phi}_{r_{0}},\mathbf{z}^{(1)}\circ\cdots\circ \mathbf{z}^{(V)}\circ \mathbf{e}_{c}\big\rangle
\end{align}
\end{footnotesize}

Because $e_{c,s}=1$ only when $c=s$ and according to Eq. (\ref{eq:xy}), we can further rewrite the equation above into
\begin{small}
\begin{align}\label{eq:inference}
& \left \langle \mathcal{W},\mathcal{Z}_{c} \right\rangle = \sum_{r_{0}=1}^{R_{0}}\cdots \sum_{r_{V}=1}^{R_{V}}g_{r_{0},\dots, r_{V}} \phi_{c,r_{0}}\big(\mathbf{z}^{(1)^{T}}\bm{\theta}_{r_{1}}^{(1)} \big)\cdots \big(\mathbf{z}^{(V)^{T}}\bm{\theta}_{r_{V}}^{(V)} \big) \nonumber \\
& = \mathcal{G}\times_{0}\bm{\phi}^{c}\times_{1}\big(\mathbf{z}^{(1)^{T}}\bm{\Theta}^{(1)} \big)\times_{2}\cdots \times_{V}\big(\mathbf{z}^{(V)^{T}}\bm{\Theta}^{(V)} \big),
\end{align}
\end{small}

where $\times_{v}$ is the $v$-mode product and $\times_{0}$ means multiplying the core tensor by the category specific vector $\bm{\phi}^{c}$.  It is worth noting that the first row $\bm{\theta}^{(v),0}$ within $\bm{\Theta}^{(v)}$ is associated with the constant value $z_{0}^{(v)}=1$ and represents the bias factors of the $v$-th view. The bias factors make the lower-order interactions active in the rating predictive function.

Using Eq. (\ref{eq:inference}) to replace the first term in Eq. (\ref{eq:pre_func}), the rating predictive function can be represented by
\begin{equation}\label{eq:final_func}
f_{c}({\mathbf{x}^{(v)}}) =\mathcal{G}\times_{0}\bm{\phi}^{c}\times_{1}\big(\mathbf{z}^{(1)^{T}}\bm{\Theta}^{(1)} \big)\times_{2}\cdots \times_{V}\big(\mathbf{z}^{(V)^{T}}\bm{\Theta}^{(V)} \big) + \mathbf{x}^{T}\mathbf{d}_{c}.
\end{equation}

\begin{figure*}[hbt]
  \centering
  \includegraphics[width=12cm]{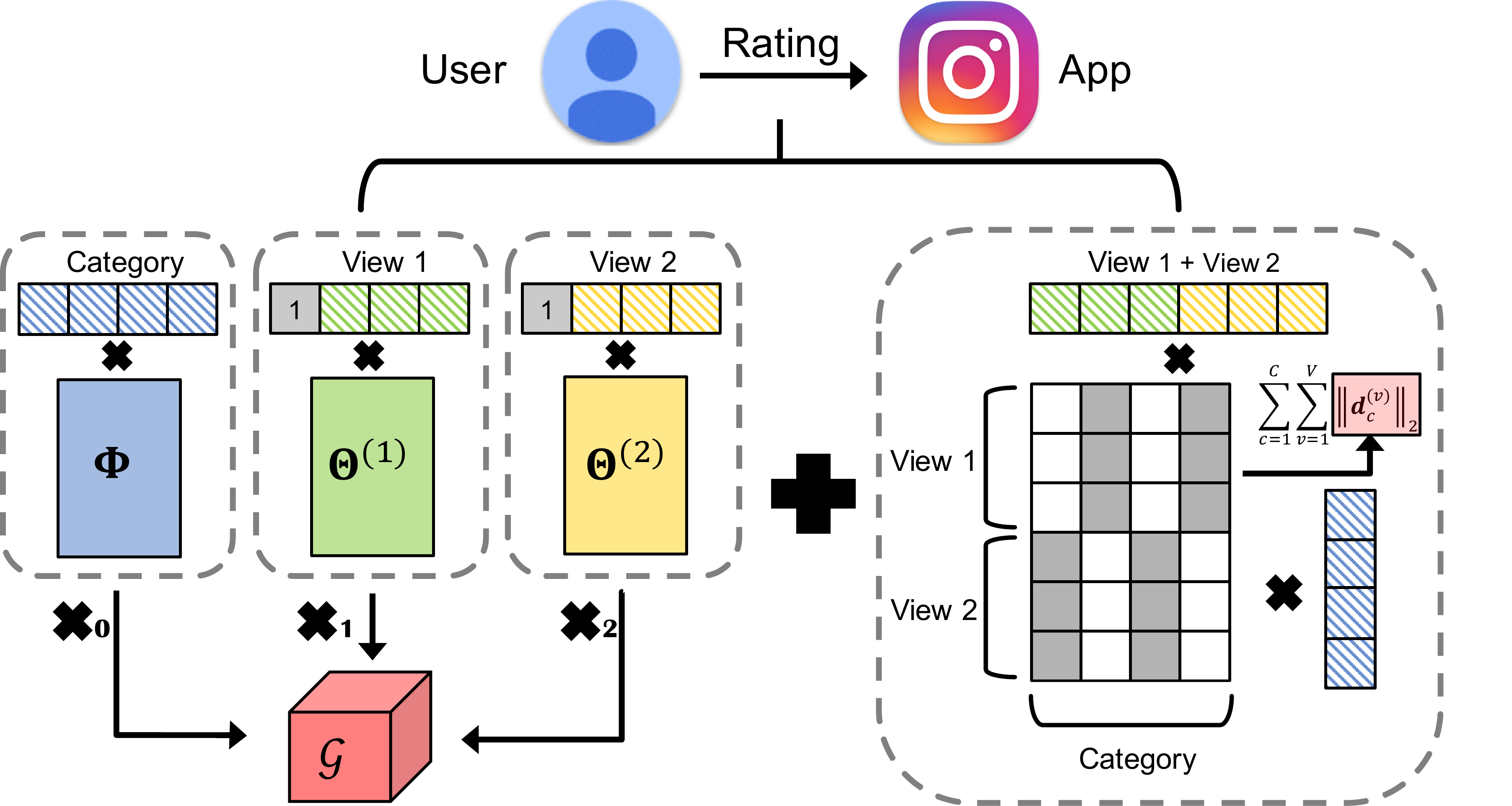}
  \caption{Framework of the proposed CATA method. The left part is the first term in Eq. (\ref{eq:final_func}) which represents the approximation of the full-order feature interactions through Tucker decomposition. The right part is the second term in Eq. (\ref{eq:final_func}) which involves the category-specific weight matrix and the group $\ell_{1}$-norm.}\label{fig:CATA}
\end{figure*}

The whole framework of the proposed CATA method is illustrated in Fig. \ref{fig:CATA}.


\subsection{Model Estimation}\label{sec:ME}
We propose to learn the app rating prediction model CATA by minimizing the following regularized empirical risk:
\begin{small}
\begin{equation}
\begin{split}
min\mathcal{H}\big(\bm{\Phi},\{\bm{\Theta}^{(v)}\},\mathcal{G}, \mathbf{D}\big) &= \sum_{c=1}^{C}\mathcal{L}_{c}\big(f_{c}\big(\{\mathbf{X}_{c}^{(v)}\}\big),\mathbf{y}_{c}\big)+\\
&\alpha\Omega_{\alpha}\big(\bm{\Phi},\{\bm{\Theta}^{(v)}\},\mathcal{G}\big)+\beta\Omega_{\beta}\big(\mathbf{D}\big).
\end{split}
\end{equation}
\end{small}

The regularization term $\Omega_{\alpha}$ and $\Omega_{\beta}$ can be set as Frobenius norm, $\ell_{2,1}$ norm, or other structural regularization. In this paper, we adopt the alternating block coordinate descent approach for the optimization of the given objective function. The whole learning procedure is summarized in Algorithm \ref{algo:CATA}.

With all other parameters fixed, the minimization over $\bm{\Theta}^{(v)}$ consists of learning the parameters $\bm{\Theta}^{(v)}$ by a regularization method, and  the partial derivative of $\mathcal{H}$ w.r.t. $\bm{\Theta}^{(v)}$ is given by
\begin{equation}
\frac{\partial\mathcal{H}}{\partial \bm{\Theta}^{(v)}} = \sum_{c=1}^{C}\frac{\partial\mathcal{L}_{c}}{\partial f_{c}}\frac{\partial f_{c}}{\partial \bm{\Theta}^{(v)}} + \alpha\frac{\partial \Omega_{\alpha}\big(\bm{\Theta}^{(v)}\big)}{\partial\bm{ \Theta}^{(v)}}
\end{equation}
where $\frac{\partial\mathcal{L}_{c}}{\partial f_{c}} = \frac{1}{N_{c}}\left[ \frac{\partial \ell_{c,1}}{\partial f_{c}},\cdots,\frac{\partial \ell_{c,N_{c}}}{\partial f_{c}}\right]^{T}\in \mathbb{R}^{N_{c}}$ and $\frac{\partial \ell_{c,n}}{\partial f_{c}} = 2(f_{c}-\mathbf{y}_{c,n})$ for $n\in [1:N_{c}]$.

For convenience, we let $\bm{\pi} \in \mathbb{R}^{1\times (R_{1}\cdots R_{V})}$ denote the Kronecker product in a reverse order from $V$ to $1$ $\prod_{v=V}^{1}\otimes \big(\mathbf{z}^{(v)^{T}}\bm{\Theta}^{(v)} \big)$ and $\bm{\pi}^{(-v)} \in \mathbb{R}^{1\times(R_{1}\cdots R_{v-1}R_{v+1}\cdots R_{V})}$ denote $\prod_{v'=V, v'\neq v}^{1}\otimes \big(\mathbf{z}^{(v')^{T}}\bm{\Theta}^{(v')} \big)$. Let $\bm{\Pi}=[\bm{\pi}_{1},\cdots,\bm{\pi}_{N}]^{T}$ and $\bm{\Pi}^{(-v)}=[\bm{\pi}_{1}^{(-v)},\cdots,\bm{\pi}_{N}^{(-v)}]^{T}$. Then we have that 

\begin{equation}\label{derivation:theta}
\begin{split}
\frac{\partial\mathcal{L}_{c}}{\partial f_{c}}\frac{\partial f_{c}}{\partial \bm{\Theta}^{(v)}}
& = \frac{1}{N_{c}}\sum_{n=1}^{N_{c}} \frac{\partial \ell_{c,n}}{\partial f_{c,n}}\frac{\partial f_{c,n}}{\partial \bm{\Theta}^{(v)}}\\
& = \sum_{n=1}^{N_{c}} \mathbf{z}^{(v)}_{c,n}\big(\frac{1}{N_{c}}\frac{\partial \ell_{c,n}}{\partial f_{c,n}}\big)\big( \bm{\pi}^{(-v)}\otimes \bm{\phi}^{c} \big) \mathbf{G}_{(v)}^{T}\\
& = \mathbf{Z}_{c}^{(v)}\big(\big(\bm{\Pi}_{c}^{(-v)}\big)^{T}\odot \big(\frac{\partial\mathcal{L}_{c}}{\partial f_{c}}\bm{\phi}^{c}\big)^{T}\big)^{T}\mathbf{G}_{(v)}^{T},
\end{split}
\end{equation}
where $\mathbf{G}_{(n)}$ the $n$-mode matricization of tensor $\mathcal{G}$.

With all other parameters fixed, the minimization over $\bm{\Phi}$ consists of learning each parameter component $\bm{\phi}^{c}$ independently. The partial derivative of $\mathcal{H}$ w.r.t. $\bm{\Phi}$ is given by
\begin{equation}
\frac{\partial\mathcal{H}}{\partial \bm{\Phi}} = \left[ \frac{\partial\mathcal{L}_{1}}{\partial f_{1}}\frac{\partial f_{1}}{\partial \bm{\phi}^{1}};\cdots ; \frac{\partial\mathcal{L}_{C}}{\partial f_{C}}\frac{\partial f_{C}}{\partial \bm{\phi}^{C}}\right] + \alpha\frac{\partial \Omega_{\alpha}(\bm{\Phi})}{\partial \bm{\Phi}}.
\end{equation}
Following the derivation in Eq.(\ref{derivation:theta}), we have that
\begin{equation}
\begin{split}
\frac{\partial\mathcal{L}_{c}}{\partial f_{c}}\frac{\partial f_{c}}{\partial \bm{\phi^{c}}}
& = \big(\frac{\partial\mathcal{L}_{c}}{\partial f_{c}}\big)^{T}\bm{\Pi}_{c}\mathbf{G}_{(0)}^{T}
\end{split}
\end{equation}

By keeping all other parameters fixed, we can get the partial derivative of $\mathcal{H}$ w.r.t. the core tensor $\mathcal{G}$ as follows,
\begin{equation}
\frac{\partial\mathcal{H}}{\partial \mathcal{G}} = \sum_{c=1}^{C}\frac{\partial\mathcal{L}_{c}}{\partial f_{c}}\frac{\partial f_{c}}{\partial \mathcal{G}} + \alpha\frac{\partial \Omega_{\alpha}(\mathcal{G})}{\partial \mathcal{G}}
\end{equation}
Following the derivation in Eq.(\ref{derivation:theta}), we have that
\begin{equation}
\begin{split}
\big(\frac{\partial\mathcal{L}_{c}}{\partial f_{c}}\frac{\partial f_{c}}{\partial \mathcal{G}}\big)_{(1\times (R_{0}\cdots R_{V}))}
& = \sum_{n=1}^{N_{c}} \big(\frac{1}{N_{c}}\frac{\partial \ell_{c,n}}{\partial f_{c,n}}\big)\big( \bm{\pi} \otimes \bm{\phi}^{c} \big) \\
& = \big(\frac{\partial\mathcal{L}_{c}}{\partial f_{c}}\big)^{T}\big( \bm{\Pi}_{c}\otimes \bm{\phi^{c}}\big)
\end{split}
\end{equation}

By keeping all other parameters fixed, we can get the partial derivative of $\mathcal{H}$ w.r.t. the core tensor $\mathbf{D}$ as follows,

\begin{equation}
\frac{\partial\mathcal{H}}{\partial \mathbf{D}} = \left[ \mathbf{X}_{1}\frac{\partial\mathcal{L}_{1}}{\partial f_{1}};\cdots ; \mathbf{X}_{C}\frac{\partial\mathcal{L}_{C}}{\partial f_{C}}\right] + \beta\frac{\partial \Omega_{\beta}(\mathbf{D})}{\partial \mathbf{D}},
\end{equation}
where $\mathbf{X}_{c}=[\mathbf{X}_{c}^{(1)};\cdots;\mathbf{X}_{c}^{(V)}]\in\mathbb{R}^{I\times N_{c}}$ is the concatenated feature matrix for the $c$-th category.


\begin{algorithm}\label{algo}
\LinesNumbered \SetAlgoVlined 
\caption{Learning CATA Model}
\label{algo:CATA}
    \KwIn{Training data $\mathcal{D}$, 
number of factors $R$, regularization parameter $\alpha, \beta$, and learning rate $\eta$}
    \KwOut{Model parameters $\{\bm{\Theta}^{(v)}\}$, $\bm{\Phi}$, $\mathcal{G}$, $\mathbf{D}$}
    \BlankLine
	Initialize$\{\bm{\Theta}^{(v)}\}$, $\bm{\Phi}$, $\mathcal{G}$, $\mathbf{D}$ $\sim\mathcal{N}(0,\sigma)$.\\ 
	\Repeat{convergence}{
		Fixing  $\{\bm{\Theta}^{(v)}\}$, $\mathcal{G}$, and $\mathbf{D}$, update $\bm{\Phi}$\\
		\For{$v=1:V$}{
			Fixing $\{\bm{\Theta}^{(v')}\}_{v'\neq v}$, $\bm{\Phi}$, $\mathcal{G}$, and $\mathbf{D}$, update $\bm{\Theta}^{(v)}$\\
		}
			Fixing  $\{\bm{\Theta}^{(v)}\}$, $\bm{\Phi}$, and $\mathbf{D}$, update $\mathcal{G}$\\
			Fixing  $\{\bm{\Theta}^{(v)}\}$, $\bm{\Phi}$, and  $\mathcal{G}$, update $\mathbf{D}$
			
	}
\end{algorithm}

\subsection{Group $\ell_{1}$-Norm}
As mentioned in section \ref{sec:ME}, the regularization terms can be Frobenius norm, $\ell_{2,1}$ norm, or other structural regularization, here we present a proper regularization term for parameter $\mathbf{D}$ to further improve the performance of rating prediction .

For the original feature spaces, \emph{i.e.,} the second term in Eq. (\ref{eq:pre_func}), the feature of a specific view might be more or less discriminative for different app categories. For instance, the description information is more useful for the distinguishing of apps in the \emph{Lifestyle} category than that of apps in \emph{Map \& Navigation} category. It is mainly because \emph{Lifestyle} is a broad cluster and the functionality which could be extracted from the description text of each app in it are very different from each other. 
Consider this, we introduce group $\ell_{1}$-norm ($G_{1}$-norm, for short) for regularization, which is defined as $||\mathbf{D}||_{G_1} = \sum_{c=1}^{C}\sum_{v=1}^{V}||\mathbf{d}_{c}^{(v)}||_{2}$ \cite{wang2012high}. The $G_{1}$-norm applies $\ell_{2}$-norm within each view and $\ell_{1}$-norm between views, so it can enforce the sparsity between different views. It means that if a specific view of features are not significant for the apps in a certain category, the weights with very small values will be assigned to them for the corresponding category.  The $G_{1}$-norm can further improve the performance of app rating prediction as it captures the global relationships between views. The right part of Fig. \ref{fig:CATA} simply shows the category-specific weight matrix as an illustration. The elements with gray color have large values. It can be found that the $G_{1}$-norm effectively emphasizes the view-wise weight learning corresponding to each category.
\section{Experiments} \label{sec:exp}
In this section, we will verify the effectiveness of the proposed method by conducting a series experiments compared to five well known baselines.


\begin{table*}[t]
  \centering
  \caption{Performance comparison on the Google Play dataset. The best results are listed in bold.}
    \begin{tabular}{c|c|c|c|c|c|c|c|c}
    \hline    
    Training & Metrics & PMF   & MTFL    & FM    & MVM   & MFM   & CATA  & CATA-G \\
    \hline
    \multirow{2}[2]{*}{60\%} & MAE   & 0.9597$\pm$0.0157 & 0.9272$\pm$0.0247 & 0.8964$\pm$0.0131 & 0.8735$\pm$0.0299 & 0.8761$\pm$0.0177 & \textbf{0.7650$\pm$0.0571} & 0.7679$\pm$0.0152 \\
          & RMSE  & 1.3015$\pm$0.0234 & 1.4616$\pm$0.0305 & 1.2133$\pm$0.0206 & 1.2122$\pm$0.0353 & 1.2021$\pm$0.0171 & 1.1720$\pm$0.0402 & \textbf{1.1586$\pm$0.0209} \\
    \hline
    \multirow{2}[2]{*}{70\%} & MAE   & 0.9463$\pm$0.0083 & 0.8889$\pm$0.0053 & 0.8786$\pm$0.0065 & 0.8496$\pm$0.0070 & 0.8660$\pm$0.0350 & 0.7911$\pm$0.0442 & \textbf{0.7864$\pm$0.0141} \\
          & RMSE  & 1.2834$\pm$0.0090 & 1.4257$\pm$0.0117 & 1.1981$\pm$0.0096 & 1.1836$\pm$0.0089 & 1.1959$\pm$0.0122 & 1.1656$\pm$0.0151 & \textbf{1.1626$\pm$0.0115} \\
    \hline
    \multirow{2}[2]{*}{80\%} & MAE   & 0.9369$\pm$0.0108 & 0.8575$\pm$0.0201 & 0.8568$\pm$0.0112 & 0.8384$\pm$0.0134 & 0.8439$\pm$0.0266 & 0.7826$\pm$0.0274 & \textbf{0.7765$\pm$0.0206} \\
          & RMSE  & 1.2745$\pm$0.0157 & 1.3904$\pm$0.0260 & 1.1785$\pm$0.0131 & 1.1741$\pm$0.0113 & 1.1815$\pm$0.0147 & 1.1502$\pm$0.0211 & \textbf{1.1419$\pm$0.0154} \\
    \bottomrule
    \end{tabular}%
  \label{tab:GP}%
\end{table*}%

\begin{table*}[t]
  \centering
  \caption{Performance comparison on the Apple App Store dataset. The best results are listed in bold.}
    \begin{tabular}{c|c|c|c|c|c|c|c|c}
    \hline
    Training & Metrics & PMF   & MTFL    & FM    & MVM   & MFM   & CATA  & CATA-G \\
    \hline
    \multirow{2}[2]{*}{60\%} & MAE   & 1.0609$\pm$0.0062 & 0.9856$\pm$0.0221 & 0.9426$\pm$0.0038 & 0.9463$\pm$0.0098 & 0.9311$\pm$0.0113 & 0.9342$\pm$0.0087 & \textbf{0.9271$\pm$0.0052} \\
          & RMSE  & 1.3180$\pm$0.0064 & 1.3064$\pm$0.0231 & 1.2890$\pm$0.0069 & 1.2717$\pm$0.0191 & 1.2422$\pm$0.0070 & 1.2396$\pm$0.0059 & \textbf{1.2377$\pm$0.0060} \\
    \hline
    \multirow{2}[2]{*}{70\%} & MAE   & 1.0551$\pm$0.0056 & 0.9856$\pm$0.0188 & 0.9345$\pm$0.0091 & 0.9409$\pm$0.0135 & \textbf{0.9198$\pm$0.0164} & 0.9246$\pm$0.0081 & 0.9247$\pm$0.0077 \\
          & RMSE  & 1.3046$\pm$0.0091 & 1.3062$\pm$0.0219 & 1.2745$\pm$0.0123 & 1.2526$\pm$0.0157 & 1.2312$\pm$0.0135 & \textbf{1.2257$\pm$0.0126} & 1.2262$\pm$0.0112 \\
    \hline
    \multirow{2}[2]{*}{80\%} & MAE   & 1.0541$\pm$0.0052 & 0.9842$\pm$0.0130 & 0.9404$\pm$0.0057 & 0.9427$\pm$0.0080 & 0.9526$\pm$0.0206 & 0.9325$\pm$0.0019 & \textbf{0.9288$\pm$0.0057} \\
          & RMSE  & 1.3065$\pm$0.0033 & 1.2955$\pm$0.0153 & 1.2800$\pm$0.0090 & 1.2493$\pm$0.0141 & 1.2372$\pm$0.0083 & 1.2317$\pm$0.0066 & \textbf{1.2286$\pm$0.0079} \\
    \bottomrule
    \end{tabular}%
  \label{tab:IOS}%
\end{table*}%

\subsection{Experimental Setup}
After the filtering for the Google Play dataset, we first select the top 20 categories with the most apps and then filter users and apps with less than 5 ratings. We obtain 3065 apps and 3895 users with 36791 rating records. The numbers of permissions and text tokens are 83 and 1762, respectively.

We randomly select $K$\% ($K=60, 70, 80$), 10\%, and 10\% of the rating records in each categories as training set, validation set, and testing set. The parameters of all the baselines are set to the optimal values. For the proposed methods, all the dimensions of the core tensor are set as 5, and the learning rate is set $\eta=0.1$. The maximum numbers of iterations are set as 400. Grid searching is employed to select the optimal regularization parameters for all the comparison methods. Each experiment is repeated for 5 times, and the mean and standard deviation of each metric in both dataset are reported in the next subsection.

We use the Mean Absolute Error (MAE) and Root Mean Square Error (RMSE) \cite{gunawardana2009survey} to evaluate the performance of the proposed approach and the other compared methods. A smaller MAE or RMSE means the better performance.

\subsection{Compared Methods}
In order to demonstrate the effectiveness of the proposed CATA approach, we compare the following methods.
\begin{itemize}
  \item \textbf{PMF.} It is the Probabilistic Matrix Factorization proposed by Salakhutdinov and Minh \cite{salakhutdinov2007probabilistic}, and the method is widely used for rating prediction tasks.
  \item \textbf{MTFL.} It is the Multi-Task Feature Learning algorithm \cite{wang2012high} which is a multivariate regression model with group $\ell_{1}$-norm. 
  \item \textbf{FM.} It is the Factorization Machine \cite{rendle2010factorization} that explores pairwise interactions between all features without view segmentation. We implement the FM by concatenating the category indicator vector and all the feature vectors as the input feature vector.
  \item \textbf{MVM.} It is the Multi-view Machine \cite{cao2016multi} that models the features from multiple views as a tensor structure to explore the full-order interactions between them. 
  \item \textbf{MFM.} It is the Multilinear Factorization Machines \cite{lu2017multilinear} that learns task-specific feature map and the task-view shared multilinear structures from full-order interactions  by applying a joint factorization.
  \item \textbf{CATA.} It is the proposed rating prediction model in this paper that effectively integrates user's preference, app category and features of multiple views and applies Tucker decomposition to learn the full-order interactions. 
  \item \textbf{CATA-G.} It is the variation of the proposed CATA that uses group $\ell_{1}$-norm for the category-specific weight matrix $\mathbf{D}$.
\end{itemize}

\subsection{Performance Comparison}
In this subsection, we present the performance comparisons between the proposed CATA methods and the baselines with respect to two metrics, \emph{i.e.,} MAE and RMSE.

Table \ref{tab:GP} and Table \ref{tab:IOS} show the performance of all the prediction methods on the Google Play and Apple App Store datasets. We can find that the proposed approach consistently outperforms the other baselines on both datasets in almost all cases. It demonstrates the superiority of the context-aware prediction approach which utilizes higher-order decomposition to learn the full-order interactions. It can be observed that CATA-G method performs better than CATA overall, which indicates that the employed group $\ell_{1}$-norm can effectively improve the rating prediction accuracy by enforcing the sparsity between different views of features.

It is not surprising that PMF has poor performance in both datasets since it doesn't employ any other features of apps. MTFL also performs badly mainly because it ignores the segmentation of feature views and the interactions between the multiple views of features. Compared to MTFL, the improvement achieved by FM illustrates the necessity of feature interactions. Both MVM and MFM outperform FM, especially for the Google Play dataset, and the results generated by them are competitive with each other. It is mainly because that MVM and MFM consider full-order interactions including the higher-order feature interactions and global bias. However, it is crucial for predicting app ratings to distinguish different categories. The propose CATA methods achieve the best performance because of the consideration of category-specific multi-view feature interactions. The application of Tucker decomposition effectively facilitates the performance as it permits the interactions within each modality \cite{cichocki2008advances} while the CP decomposition used in MFM does not. 

Comparing the two datasets, it can be found that the superiority of the proposed approach is more significant for Google Play dataset. It is mainly caused by the fewer categories and feature views in the dataset of Apple's App Store. The fewer categories might not sufficiently discriminate the important features in the specific category, while the fewer views of features would lead to the lack of some interaction information between features from different views. Moreover, the two categories in Apple App Store, \emph{i.e.,} ``Free'' and ``Paid'', do not have their own unique characteristics and are not easy to differentiate from each other. Nonetheless, even with very limited context information, the CATA methods still outperform the baselines in almost all cases.

\subsection{Impact of Feature Views}
\begin{figure}
\centering
\subfigure{\includegraphics[width=4.3cm]{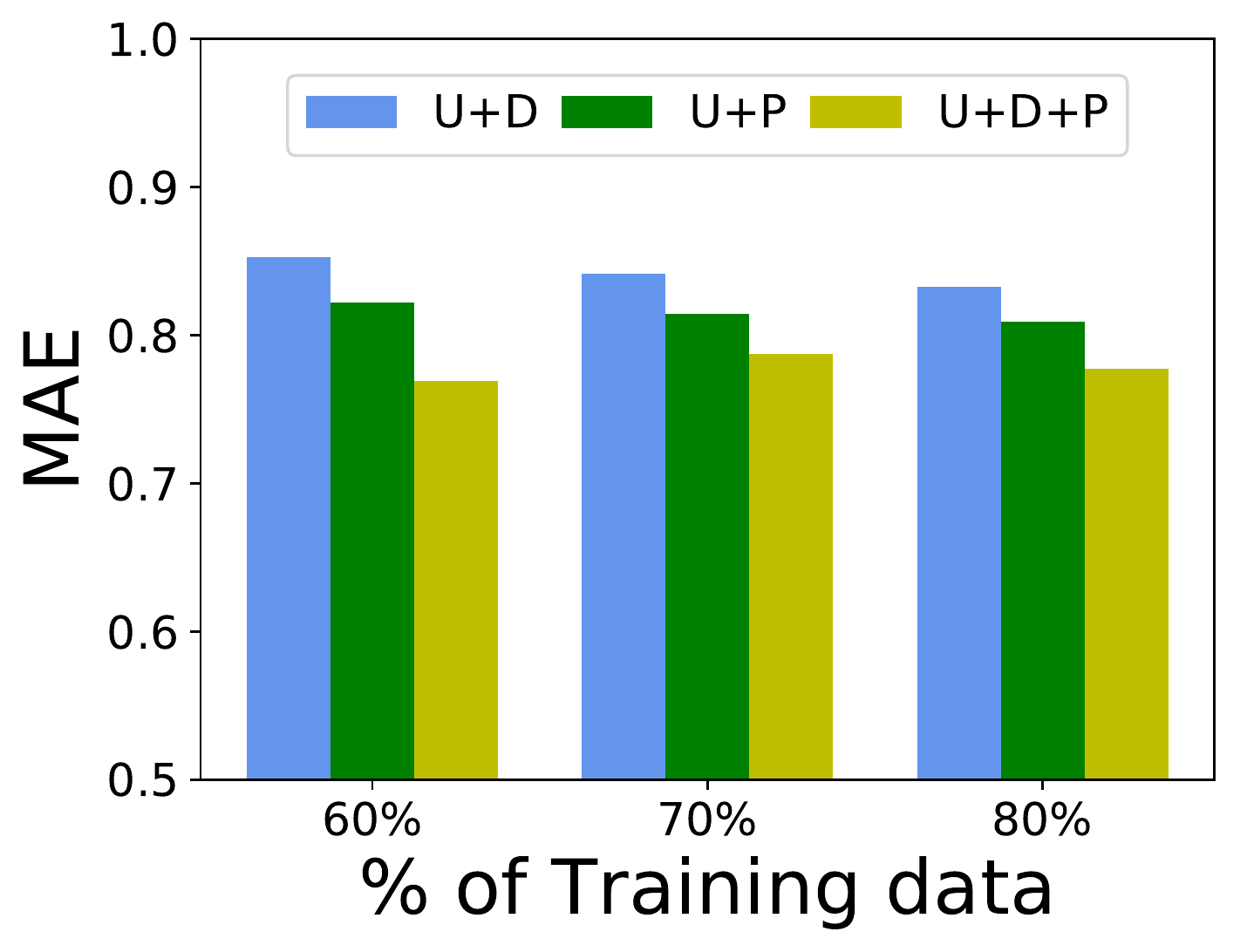}}
\subfigure{\includegraphics[width=4.3cm]{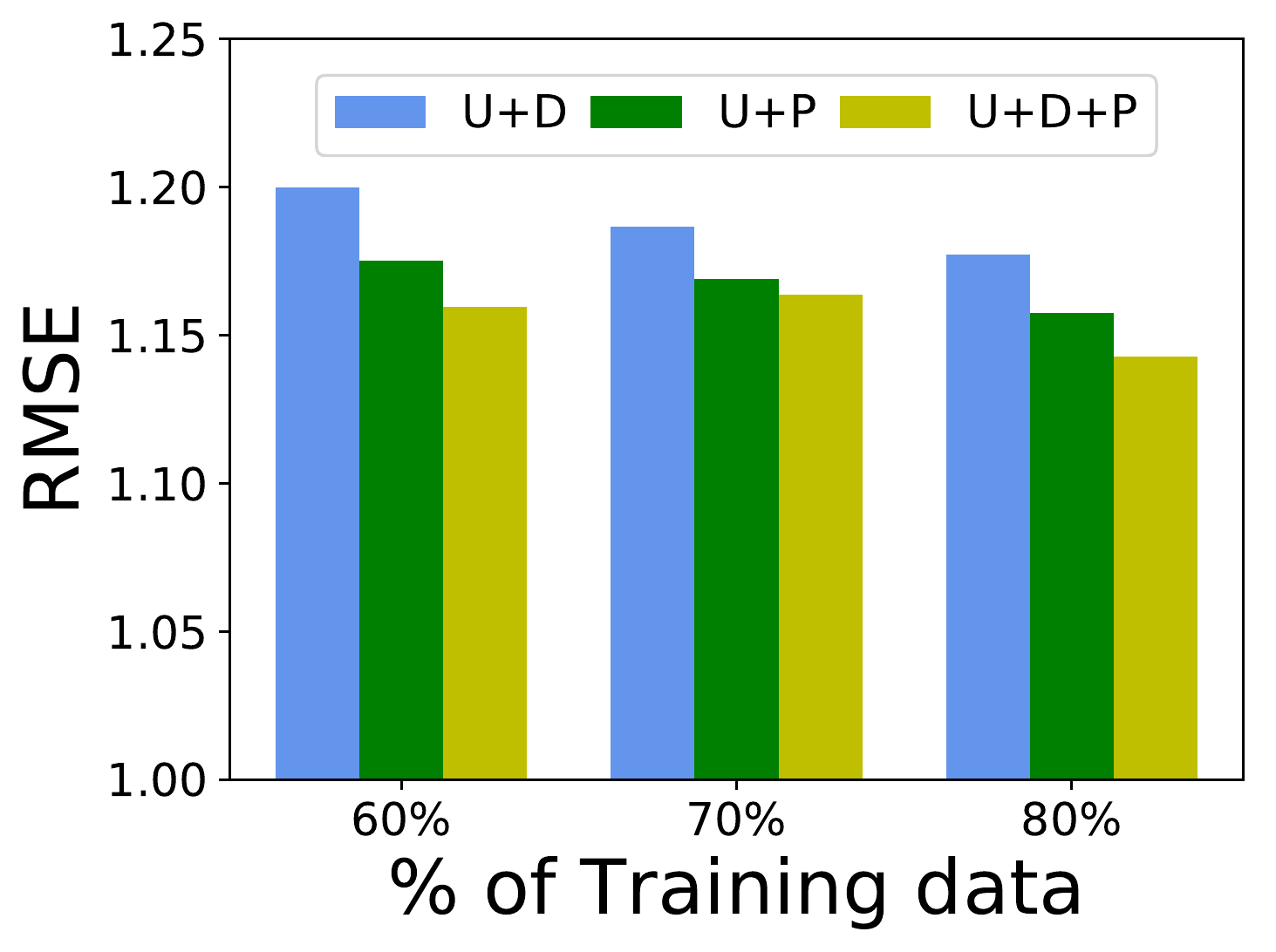}}
\caption{Performance of the CATA-G method with different feature views on Google Play dataset.}\label{fig:view_test}
\end{figure}

In order to explore the impact of the feature views for the proposed CATA-G method, we conduct experiments based on different numbers of views. As each rating record in Apple App Store dataset only has two views, \emph{i.e.,} user and review text, we use Google Play dataset for the experiments. Figure \ref{fig:view_test} shows the prediction performance of the CATA-G method with two and three feature views. Note that U, D, and P respectively denote user, description text, and permission. It can be observed that the CATA-G method consistently performs best when incorporating three views of features, which benefits from the complementary information generated by the interactions among the various views of features. It indicates that the incorporation of multiple views of features can effectively improve the accuracy of rating prediction for apps. Consider the results produced by two views, we can find that the adoption of permission brings better results than that of description text. This is probably because the features extracted from description text are more sophisticated and higher dimensional, which may provide redundant information. 
 
\subsection{Category-Specific Performance}
\begin{figure}
\centering
\subfigure{\includegraphics[width=4.3cm]{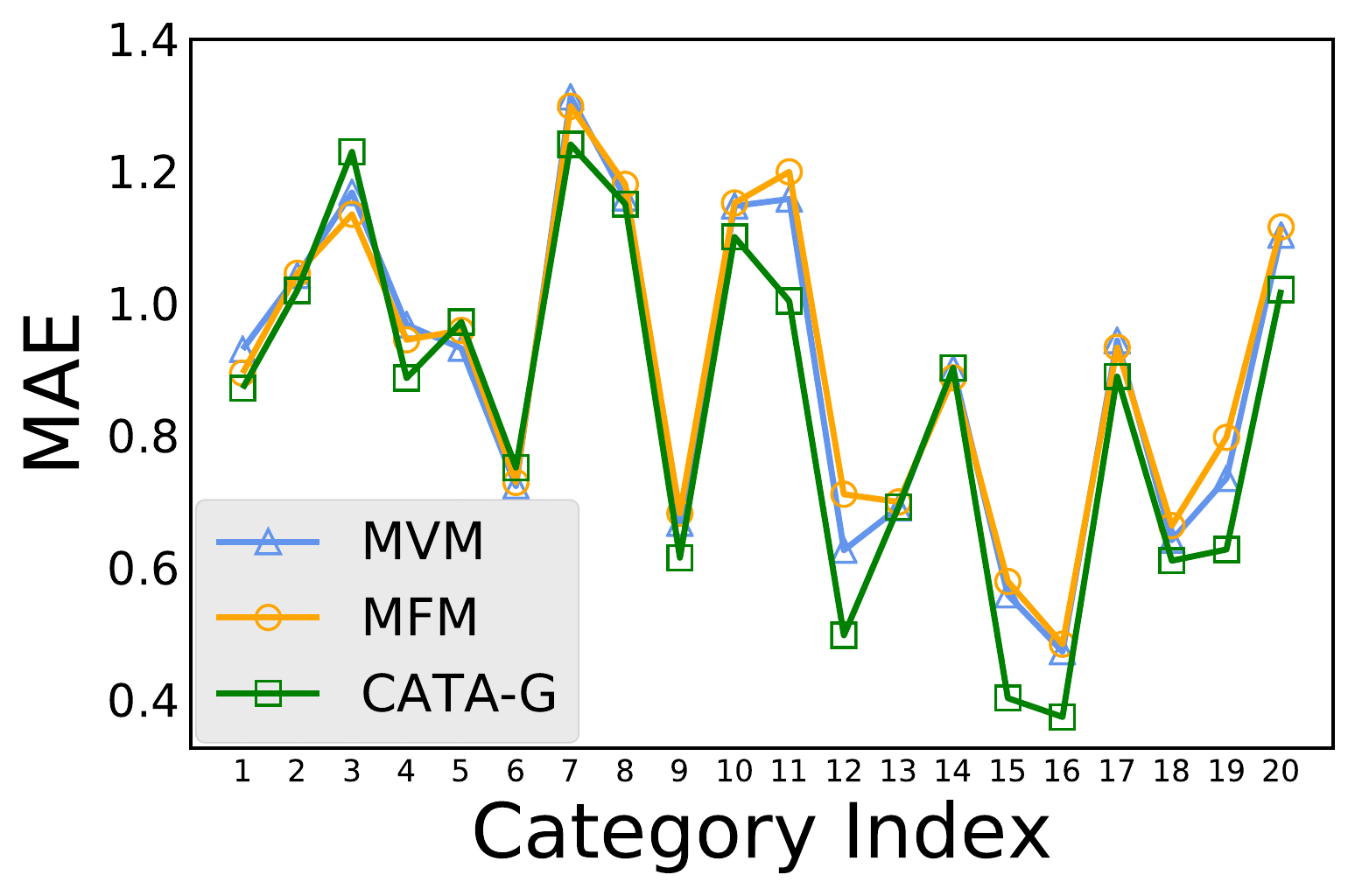}}
\subfigure{\includegraphics[width=4.3cm]{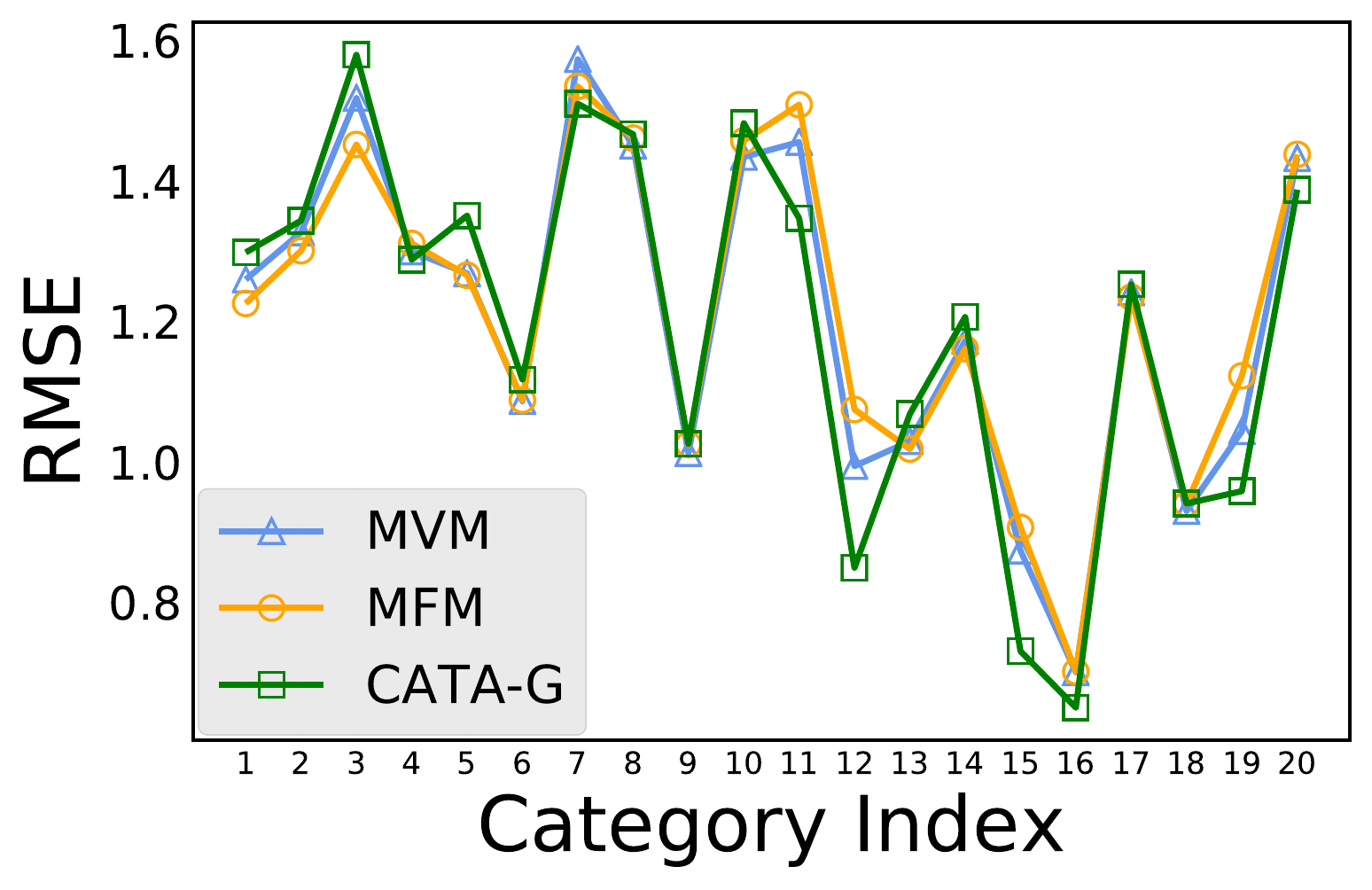}}
\caption{Category-specific performance of MVM, MFM, and CATA-G on Google Play dataset. The category indexes on the $x$-axis is sorted by the numbers of rating records within the categories in an ascending order.}
\label{fig:cate_performance}
\end{figure}

In this section, we further analyze the performance of the proposed method for each category based on Google Play dataset. Figure \ref{fig:cate_performance} shows the MAE and RMSE values of the proposed method and the top 2 baseline methods in each category. The category indexes on the $x$-axis is sorted by the numbers of rating records within the categories in an ascending order. We can find that MFM performs better than MVM in the categories with few training instances. It indicates that when few instances are available, the method incorporating the category information can improve the performance as it explores the information from other complementary information. The performance of the proposed CATA-G method is the worst with few instances, as CATA-G has more model parameters to learn and requires more instances. For the categories with more instances, CATA-G makes significant improvements and outperforms the other two methods. Compared with MFM, the superiority of CATA-G is the application of Tucker decomposition which can effectively retain the principal components of the weight tensor. Another intereting observation in Fig. \ref{fig:cate_performance} is that CATA-G makes the top 5 improvements for category \emph{Simulation}, \emph{Action}, \emph{Casual}, \emph{Arcade}, and \emph{Puzzle} (\emph{i.e.,} \#11, \#12, \#15, \#16, and \#19). The apps in the five cateogries are game apps, and it means the features of them are more complicated as each game app has its specific theme setting. Therefore, the proposed method has a greater ability to discriminate the importance of each feature in a complicated feature sets. 

\subsection{Sensitivity Analysis}
\begin{figure}
\centering
\subfigure[]{\includegraphics[width=4.3cm]{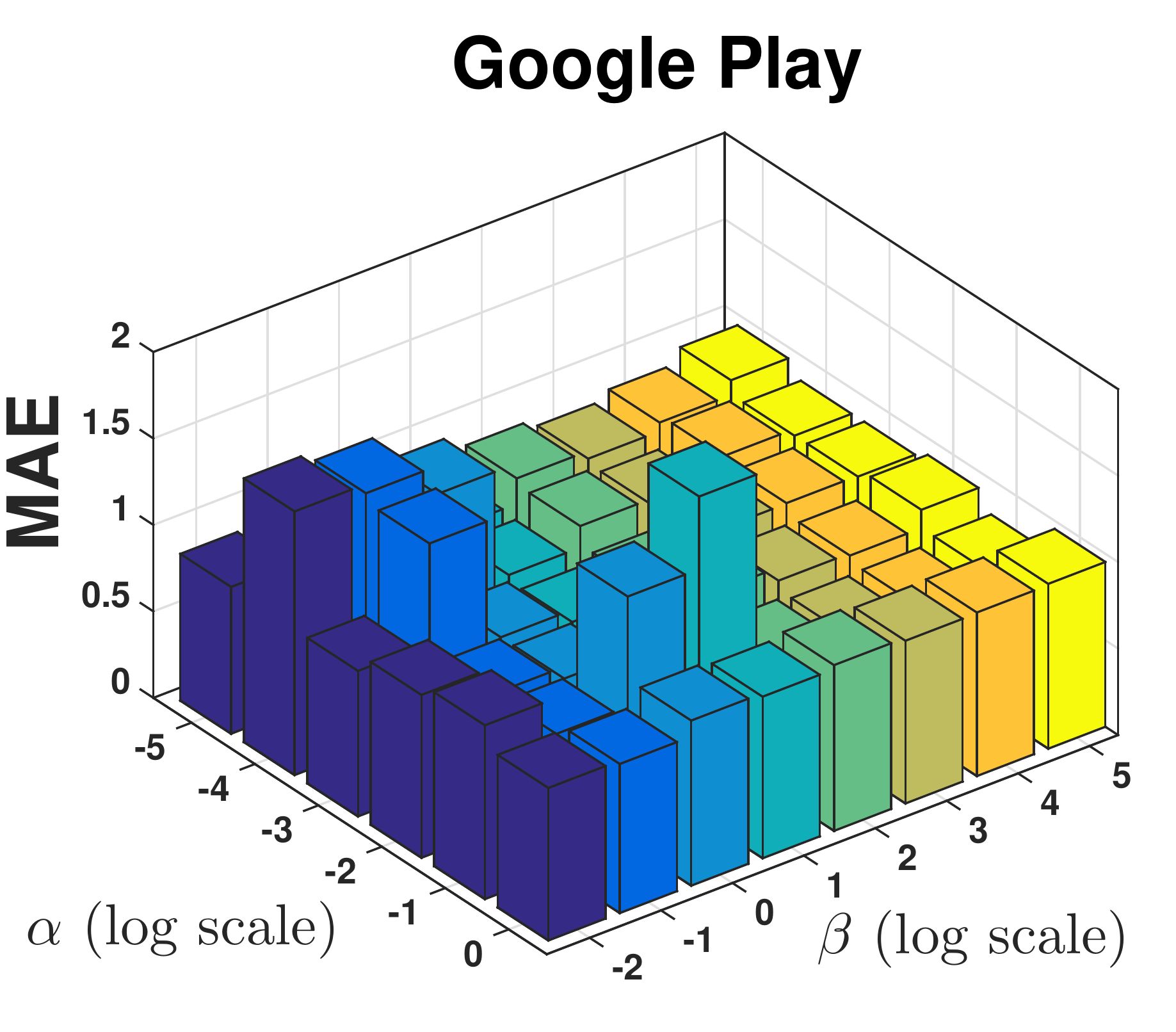}}
\subfigure[]{\includegraphics[width=4.3cm]{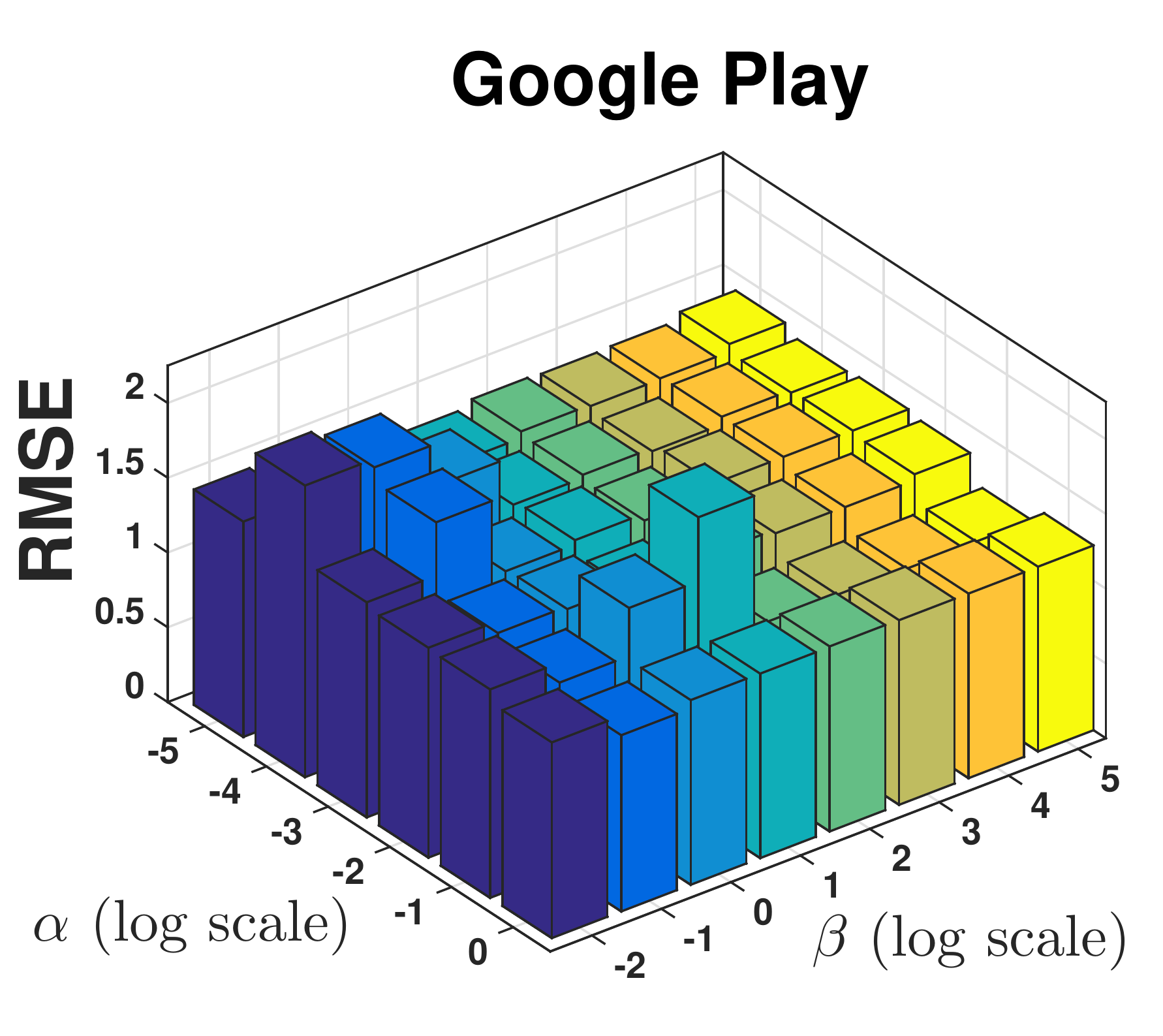}}
\subfigure[]{\includegraphics[width=4.3cm]{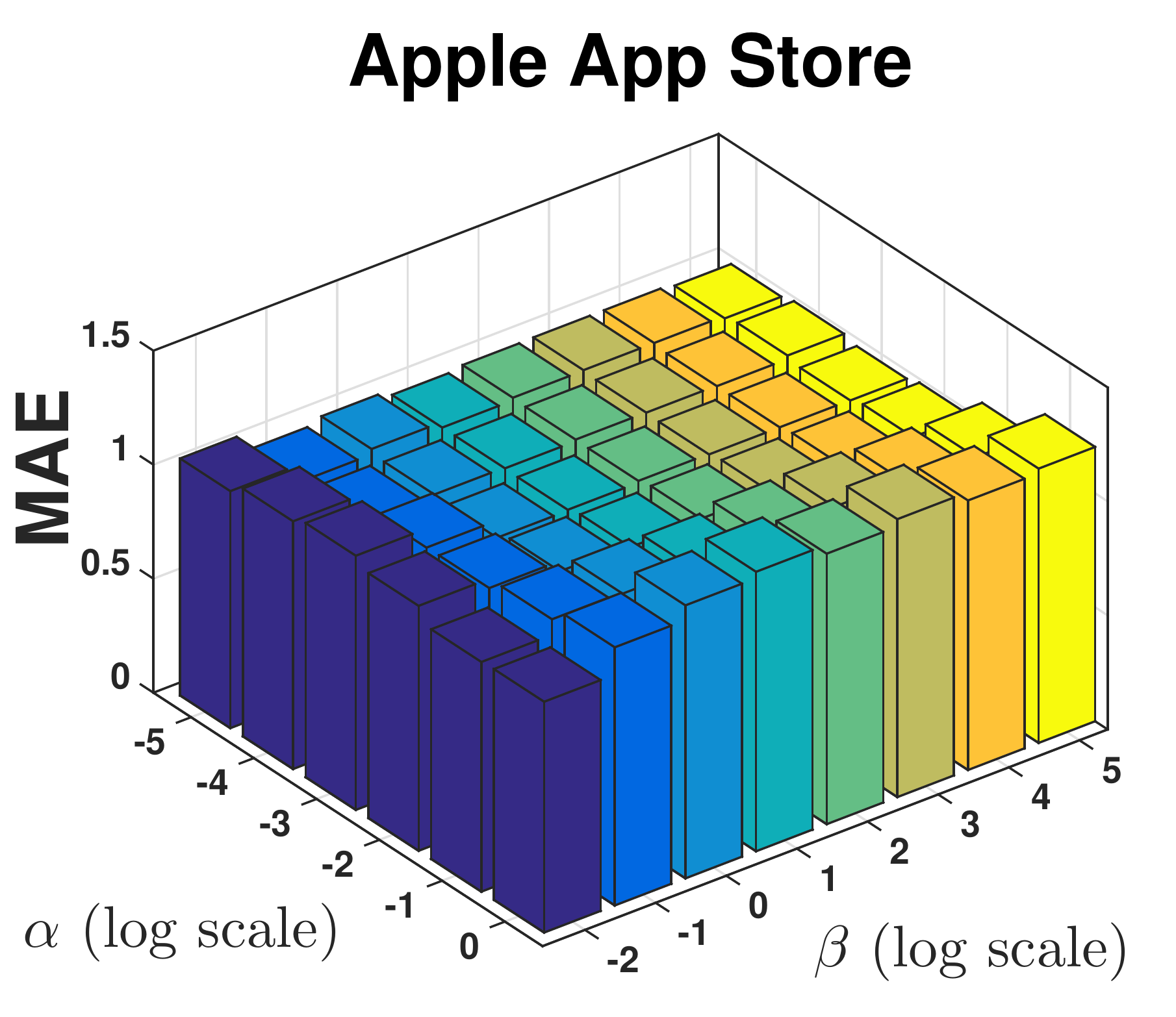}}
\subfigure[]{\includegraphics[width=4.3cm]{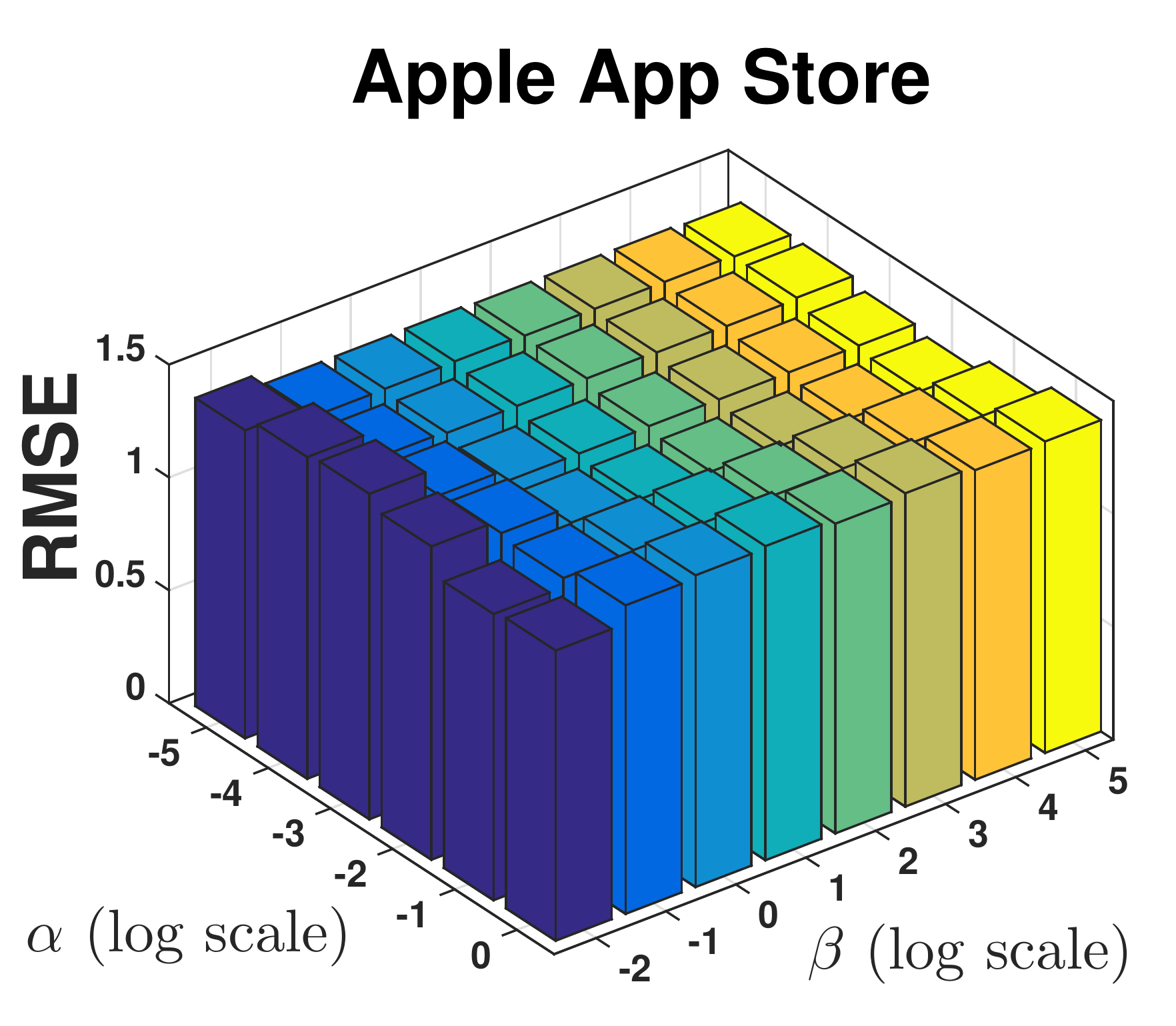}}

\caption{Sensitivity analysis of hyper-parameters.}
\label{fig:para_analysis}
\end{figure}

There are two hyper-parameters (\emph{i.e.,} $\alpha$ and $\beta$) in the proposed CATA approach. They are used to control the trade-off between the empirical loss and the prior knowledge encoded by the regularizations. To learn the impacts of the two hyper-parameters on the performance of app rating prediction, we run the proposed approach with different values for $\alpha$ and $\beta$ on the two datasets. 
From Fig. \ref{fig:para_analysis}, we can observe that the performance is stable for most pairs of the two hyper-parameters. For each dataset, the effects of the two hyper-parameters on MAE and RMSE are different. For the Google Play dataset, Figs. \ref{fig:para_analysis}(a) and (b) show that the unstable and worse MAE and RMSE are produced when given a larger $\alpha$ (\emph{i.e.,} $\alpha=1$) or a smaller $\beta$ (\emph{i.e.,} in the range from $10^{-2}$ to $10$). And the best performance is achieved by the relatively large value of $\beta$ (\emph{i.e.,} in the range from  $10^{2}$ to $10^{5}$) with $\alpha = 0.1$. Figure \ref{fig:para_analysis} (c) and (d) report the results of the Apple App Store dataset, from which we can find that the performance is more stable than that of Google Play. Similarly, when the value of $\alpha$ is larger or the value of $\beta$ is smaller, the MAE and RMSE are relatively higher. The best performance of the MAE is achieved by $\alpha = 10^{-3}$ while the RMSE is much lower when the value of $\alpha$ is set $10^{-2}$.  The value of $\beta$ is in the range from $10^{0}$ to $10^{5}$. For both datasets, the best performance is generated by a larger $\beta$ and the larger $\beta$ means the model hyper-parameters for the category-specific weight matrix $\mathbf{D}$ can be small. It indicates that the part of full-order interactions among multiple categories and multiple views of features is much more important.

\section{Related Work}\label{sec:related}
To the best of our knowledge, this is the first work to consider mining the full-order interactions among app context information with tensor analysis to facilitate mobile app recommendation. From the conceptual perspective, two topics can be seen as closely related to this work: mobile app recommendation, tensor factorization and its applications. We give a short overview of these areas and distinguish from other existing methods.

\textbf{Mobile App Recommendation} has drawn an increaasing number of attentions as an effective way to alleviate information overload in app market. Most of the existing works are trying to leverage one or more kinds of features to improve the recommendation performance. Yan \emph{et al.} \cite{yan2011appjoy} developed the AppJoy system that recommends mobile apps by based on the analysis of users' usage records. In \cite{yin2013app}, Yin \emph{et al.} applies users' view/download sequences to mine the actual value and tempting value of apps, which are used to build a recommendation model considering the contest between apps. Features from the other sources are incorporated in some works. For instance, to address the cold-start problem, Lin \emph{et al.} \cite{lin2013addressing} proposed to apply app followers' features collected from Twitter to model the app and estimate which users may like the app. 
Zhu \emph{et al.} \cite{zhu2014mobile} presented a method to evaluate the security risks of apps and proposed a flexible app recommendation approach combining both apps' popularity and users' security preferences through the modern portfolio theory. A recommendation model which can capture the trade-off between app functionality and user privacy preference was proposed in \cite{liu2015personalized}. However, most of these works did not consider the complex interactions among the features of different views. In this work, we propose to model the interactions as a tensor structure and leverage tensor factorization to learn the latent relationships.

\textbf{Tensor Factorization and Applications} Tensor factorization is a method to divide a tensor in multidimensionality into many smaller parts. A comprehensive survey on tensor factorization can be found in~\cite{kolda2009tensor}. Two well-known methods in this area are CANDECOMP/PARAFAC (CP) factorization and Tucker factorization. Both of them can be considered as higher-order generalization of Singular value decomposition (SVD) and Principle Component Analysis (PCA). These methods are used to decompose tensor data into simpler form, containing better features and intrinsic multi-way structures. CP factorization has been frequently investigated in the multi-view learning literature because of its simplicity. Specifically, \cite{cao2014tensor} first introduced to use the outer product operator to fuse multi-view features in the tensor structure and proposed a CP factorization based multi-view feature selection method. Later, \cite{cao2016multi} extended this approach to consider the full-order interactions between features, and proposed a CP factorization based multi-view machine (MVM) for multi-view prediction problems. Recently, \cite{lu2017multilinear} extended the MVM method to deal with multi-task multi-view prediction problems and proposed a CP factorization based multilinear factorization machine. However, to the best of our knowledge, none of the studies explored Tucker decomposition in the scenario of multi-view learning. Tucker decomposition is more general than the CP decomposition, and permits the interactions within each mode while the CP decomposition does not \cite{cichocki2008advances}. This paper gives an application of Tucker decomposition into multi-view learning task.





\section{Conclusions} \label{sec:conclude}
In this paper, we propose a context-aware recommendation approach based on tensor analysis (CATA) for mobile apps. The proposed CATA approach models the interactions among the multiple categories and multiple views of features of apps as a tensor structure. CATA applies the Tucker decomposition to collectively learn the category-specific features and the latent relationships integrated in the full-order interactions without physically building the tensor. To further improve the performance of app recommendation, we present a group $\ell_{1}$ norm regularization for the global category-specific weight matrix.
Extensive experiments based on two real-world app datasets demonstrate the effectiveness of the proposed CATA approach.



\bibliographystyle{IEEEtran}


\end{document}